\documentclass[conference]{IEEEtran}

\usepackage{cite}
\usepackage{graphicx}
\usepackage{xspace}
\usepackage{xcolor}
\usepackage{array}
\usepackage{booktabs}
\usepackage{tabularx}
\usepackage{multirow}
\usepackage{algpseudocode}
\usepackage{url}

\graphicspath{{assets/}{./}}

\newcommand{\sys}{SPECA\xspace}

\providecommand{\Description}[1]{}

\begin{document}

\title{Beyond Code Reasoning: Specification-Anchored Auditing of Multi-Implementation Distributed Protocols}

\author{%
\IEEEauthorblockN{Masato Kamba}
\IEEEauthorblockA{Nyx Foundation\\
masato.kamba@nyx.foundation}
\and
\IEEEauthorblockN{Hirotake Murakami}
\IEEEauthorblockA{Aichi Prefectural Aichi High School\\of Technology and Engineering\\
2009hirotake@gmail.com}
\and
\IEEEauthorblockN{Akiyoshi Sannai}
\IEEEauthorblockA{Department of Physics, Kyoto University\\
sannai.akiyoshi.7z@kyoto-u.ac.jp}
}

\maketitle

\begin{abstract}
Code-driven auditing succeeds where vulnerabilities surface as
code-level anomalies, but fails when correctness depends on what the
specification \emph{requires} rather than how the code is written.
Production blockchain networks expose this gap directly: byzantine
consensus is reached by running many independent clients of a shared
specification, so a specification-divergence defect in any one client
can fork the network, halt finality, or silently corrupt cross-client
state. Existing auditing tools reason one repository at a time and
have no shared baseline that can be held constant while the
implementation varies.

We present \sys, an LLM-driven audit framework that derives explicit,
categorized security properties (invariants, preconditions,
postconditions, trust assumptions) from natural-language specifications
and reuses them across implementations. \sys yields three capabilities
absent from code-driven auditing: controlled cross-implementation
comparison under a constant property set, detections grounded in
specification invariants no code-level pattern encodes, and false
positives traceable to a specific pipeline phase rather than opaque
model errors.

We evaluate on two benchmarks. On the Sherlock Ethereum Fusaka Audit
Contest (10 targets, 366 auditor submissions), a post-contest
re-experiment recovers all 15 in-scope H/M/L vulnerabilities
expert-augmented (8/15 automated-only) and independently surfaces 4
fix-confirmed bugs, including a cryptographic-invariant violation
missed by every adjudicated contest finding; \emph{all 15} detecting
properties encode specification-derived information unavailable from
code alone. On the RepoAudit C/C++ benchmark, \sys reaches 88.9\%
precision at 100\% recall ($F_1{=}0.94$, on par with RepoAudit's best
configuration) and surfaces 12 author-validated bugs beyond ground
truth, with two externally validated (Level~A fix; Level~B maintainer
approval). Symmetrically, \sys flags 5 of RepoAudit's 40 published
bugs as defensive-coding fixes with no reachable exploit path
(DeepSeek~R1 rejects all 5; Sonnet~4.5 rejects 4 of 5)---a verdict the
explicit property representation makes auditable. All false positives
trace to three pipeline-pinned root causes (trust boundary,
code-reading, specification misinterpretation), and a multi-model
study identifies property-generation quality, not model reasoning, as
the binding constraint on coverage. End-to-end cost is $\sim$\$30 per
H/M/L bug ($\sim$42~min wall-clock under parallel execution).
\end{abstract}


\IEEEpeerreviewmaketitle

\section{Introduction}
\label{sec:introduction}

Production blockchain networks reach byzantine consensus by running multiple independent client implementations of a shared protocol specification.
Ethereum's mainnet alone is sustained by clients in Go, Rust, Nim, TypeScript, Java, and C\#, each independently engineering the same consensus and execution rules.
This client diversity bounds the blast radius of any single bug, and it is the operational reason a specification-divergence defect in any one client is not a local code error but a \emph{distributed-systems} failure that can fork the network, halt finality, or silently corrupt cross-client state.

In a recent security contest exercising exactly this setting, 366 auditors examined ten implementations of the same cryptographic protocol specification and collectively identified fifteen vulnerabilities---all recognizable as local code defects.
Yet a subtle defect remained absent from every adjudicated finding: in a core cryptographic library, the KZG proof verification routine passed the original commitment array rather than its deduplicated variant, silently undermining the mathematical guarantee that the verification scheme depends on.
The bug was a violation of an invariant that the specification defines but no code-level auditing tool was designed to check.
This is the structural blind spot we target: when a vulnerability arises not from how code is written but from what the specification \emph{requires}, repository-local auditing has neither the vocabulary to detect it nor the structure to explain the miss.

Recent code-driven analyses---repository-level auditing agents~\cite{repoaudit}, LLM-enhanced static analyzers~\cite{llift,buglens}, and code-derived property generators~\cite{propertygpt}---all start from the code and work outward.
They succeed when vulnerabilities surface as recognizable code-level anomalies, but they cannot express security expectations that the specification defines and the code never explicitly encodes, and their false positives are opaque: when a finding is wrong, there is no principled account of \emph{why} the tool diverged from reality.

Our key insight is that auditing becomes both more effective and more analyzable when vulnerabilities are represented as violations of explicit security properties derived from the specification, rather than as repository-local code patterns.
We derive typed properties from natural-language specifications and check each via Hoare-style proof-attempt reasoning: the LLM treats the property as a per-program-point obligation and emits a finding only when no discharging argument is reachable.
This grounds every finding in a traceable security expectation and makes false positives interpretable---each maps to one of three pipeline-phase-specific root causes (trust-boundary misunderstanding, code-reading errors, specification misinterpretation), turning opaque model errors into ranked engineering targets.
Because a missed specification invariant in a production client can fork the network, we deliberately optimize \sys for \emph{recall-first} auditing: a costly false positive (manual triage time) is acceptable when the alternative is a silent consensus split, and the explicit property representation is what keeps triage tractable.

\paragraph{Scope and threat model.}
We focus on multi-implementation distributed protocols whose specifications are publicly maintained natural-language artifacts.
Ethereum's consensus and execution specifications are our primary case study; the framework targets any protocol that prescribes byzantine-fault-tolerant behavior across heterogeneous clients (e.g., Cosmos, Solana, QUIC).
The audit consumes the specification and source code, trusts the build toolchain and specification text, and models an adversarial peer that can craft protocol messages or local inputs but cannot modify the binaries under audit.
Side-channel and supply-chain attacks are out of scope.
We implement \sys on a current-generation LLM (Claude Sonnet~4.5) and report weaker models as ablations.
We evaluate on two established benchmarks: the Sherlock Ethereum Fusaka contest (10 audit targets, 366 professional auditors) and RepoAudit (15 C/C++ projects, where API documentation and module-level invariants serve as the specification surface).
The evaluation reported here is a controlled post-contest re-experiment of \sys against the adjudicated ground truth, isolating methodology from operator effects; the KZG defect is among the bugs recovered.

\sys delivers two coupled outcomes: vulnerabilities that a code-driven pipeline cannot articulate, and a structured account of why each issue was found or missed.
This paper makes the following contributions:

\begin{itemize}
\item \textbf{Cross-implementation comparison under a shared property vocabulary.} Properties derived once from the specification are reused across 10 Ethereum audit targets---7 consensus/execution clients, 1 EVM library, and 2 KZG cryptographic libraries spanning 6 languages---holding the security baseline constant while the implementation varies (\S\ref{sec:design}, \S\ref{sec:evaluation}).

\item \textbf{Specification-anchored audit framework.} \sys recovers all 15 in-scope H/M/L Sherlock vulnerabilities expert-augmented (8/15 automated-only) and independently discovers 4 fix-confirmed bugs, including a cryptographic-invariant violation missed by all 366 contest submissions; on RepoAudit it matches the best published precision (88.9\%) and surfaces 7--18 author-validated beyond-GT candidates across three models, with 2 cross-model candidates externally validated to Level~A/B (\S\ref{sec:evaluation}).

\item \textbf{Pipeline-phase-traceable failure analysis.} False positives map to three interpretable root causes pinned to specific phases; a multi-model study identifies property-generation quality as the binding constraint on coverage, turning model errors into ranked engineering targets (\S\ref{sec:evaluation}, \S\ref{sec:discussion}).

\item \textbf{Engineering principles for recall-first auditing.} We isolate the productive property types (invariants 75\% precision, preconditions 100\%) and characterize a severity-preserving review filter that removes 30 of 102 raw findings while losing zero H/M/L true positives. The full pipeline costs roughly \$30 per H/M/L bug ($\sim$42-minute wall-clock under parallel execution), giving design principles for high-assurance distributed-systems tooling (\S\ref{sec:evaluation}, \S\ref{sec:discussion}).
\end{itemize}

\section{Related Work}
\label{sec:related}

We organize related work along three axes and summarize the closest systems in Table~\ref{tab:related-summary}.
Appendix~\ref{app:related} provides a full discussion; here we highlight what distinguishes \sys.

\noindent\textbf{LLM-augmented code-level detection.}
LLift~\cite{llift} and BugLens~\cite{buglens} guide LLMs through structured reasoning for kernel use-before-initialization and taint-style bugs; IRIS~\cite{iris} and LLMxCPG~\cite{llmxcpg} combine LLMs with CodeQL or code property graphs; RepoAudit~\cite{repoaudit} deploys an autonomous repository-level agent.
All start from repository-local information (dataflow, code property graphs, path constraints).
When correctness depends on normative requirements the code never explicitly encodes, code-level starting points cannot express whether an implementation satisfies its intended security expectation~\cite{ullahsp24, primevul}.

\noindent\textbf{LLM-driven property generation.}
PropertyGPT~\cite{propertygpt} (NDSS'25 Distinguished Paper), GPTScan~\cite{gptscan}, and iAudit~\cite{iaudit} target smart contracts; FormalBench~\cite{formalbench} finds that multi-phase pipelines with structured review are necessary for reliable inference.
These share \sys's emphasis on explicit properties as the unit of analysis, but derive them primarily from code or existing properties rather than authoritative natural-language specifications, and focus predominantly on smart contracts.

\noindent\textbf{Specification-driven analysis.}
RFCAudit~\cite{rfcaudit} audits protocol implementations against RFCs with a hierarchical code-indexing agent; EDEFuzz~\cite{edefuzz} fuzzes web APIs using API specs; Jin et al.~\cite{jinase25} show that complex prompting paradoxically raises false-negative rates due to overcorrection bias---a result motivating \sys's recall-first filtering.
Even ideal specifications~\cite{cortiereprintideal} can miss intuitively expected security properties.
RFCAudit is closest in spirit but uses implicit RFC compliance as its unit of reasoning, whereas \sys derives explicit, typed security properties and supports cross-implementation comparison under a shared property vocabulary plus a structured false-positive taxonomy traced to pipeline-phase origins.
Related property-enforcement work~\cite{quack, islab} and studies of specification--implementation gaps in blockchain systems~\cite{nayakccs24} situate specification-based analysis as an active area \sys extends from enforcement to auditing.

\begin{table*}[t]
\centering
\caption{Comparative summary of approaches closest to \sys. \sys's primary differentiators are the \emph{combination} of explicit property grounding, cross-implementation comparability, and structured false-positive analysis.}
\label{tab:related-summary}
\small
\setlength{\tabcolsep}{8pt}
\begin{tabular}{l l l l l}
\toprule
\textbf{Axis} & \textbf{RepoAudit} & \textbf{PropertyGPT} & \textbf{RFCAudit} & \textbf{\sys} \\
\midrule
Source of expectations    & Code paths   & Code + properties & RFC NL     & NL spec + threat models \\
Unit of reasoning         & Bug pattern  & Formal property   & RFC clause & Typed property (inv/pre/post/asm) \\
Cross-impl.\ comparability & No           & No                & No         & Yes \\
Structured FP taxonomy    & No           & No                & No         & Yes (3 causes, phase-origin) \\
\bottomrule
\end{tabular}
\end{table*}

\section{\sys Framework Design}
\label{sec:design}

\subsection{Overview}
\label{sec:design:overview}

\begin{figure*}[t]
\centering
\includegraphics[width=0.95\textwidth]{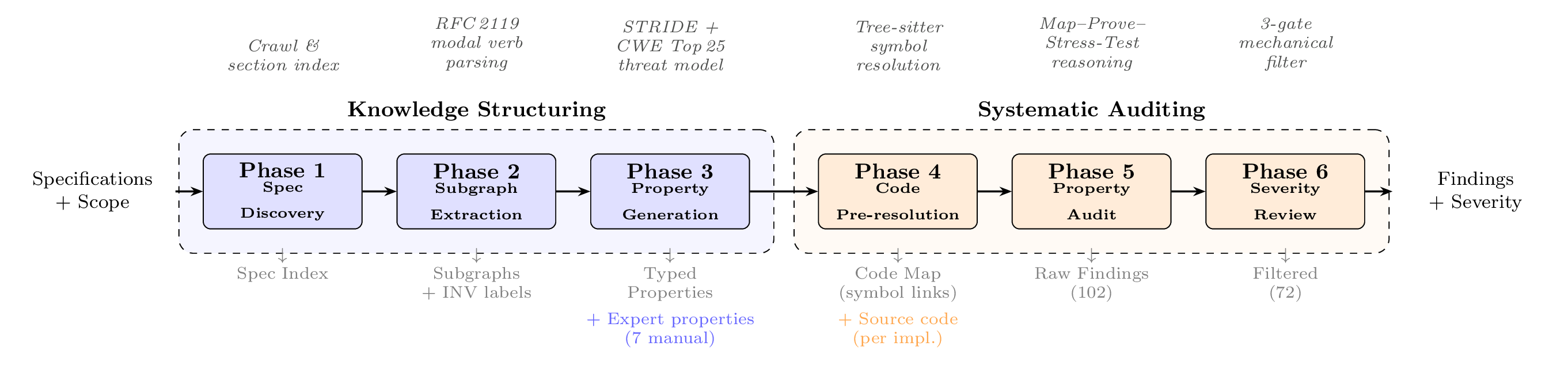}
\caption{The \sys pipeline. Phases~1--3 (\emph{Knowledge Structuring}) derive security properties from specifications; Phases~4--6 (\emph{Systematic Auditing}) verify properties against implementations. In multi-implementation settings, the left stage executes once and the right stage executes per target.}
\label{fig:pipeline}
\end{figure*}

\sys automates the workflow of a specification-aware security auditor through a six-phase pipeline organized into two stages: \emph{knowledge structuring} (Phases~1--3), which transforms natural-language specifications into explicit security properties, and \emph{systematic auditing} (Phases~4--6), which applies structured proof-attempt reasoning to check whether implementations satisfy those properties.
Figure~\ref{fig:pipeline} illustrates the end-to-end pipeline: \sys first builds a structured representation of what the code \emph{should} do (from specifications and threat models), then checks whether the code \emph{actually} does it through proof construction, grounding every finding in a traceable security property.

The workflow proceeds in six phases:
\begin{enumerate}
\item \textbf{Specification Discovery} (Phase 1): Crawl specification documents and produce a structured index.
\item \textbf{Subgraph Extraction} (Phase 2): Decompose each specification into program graphs with invariant annotations derived from RFC~2119 modal verbs, following Nielson and Nielson~\cite{nielson1992}.
\item \textbf{Property Generation} (Phase 3): Apply STRIDE threat modeling with CWE Top~25 augmentation to derive typed security properties (Table~\ref{tab:property-types}) from subgraphs and the bug bounty scope.
\item \textbf{Code Pre-resolution} (Phase 4): Link each property to concrete source locations using Tree-sitter symbol analysis, reducing downstream audit tokens by 40--60\%.
\item \textbf{Property-Grounded Audit} (Phase 5): For each property, attempt to construct a structured proof that it holds; gaps constitute findings.
\item \textbf{Severity-Preserving Review} (Phase 6): Filter false positives through three narrow mechanical gates (dead code, trust boundary, scope) designed to preserve H/M/L recall.
\end{enumerate}

In multi-implementation settings, Phases~1--3 execute once against the specification, producing a shared property vocabulary; Phases~4--6 execute independently per target. Holding security expectations constant while varying the code under audit yields a controlled comparison: the same property vocabulary produces systematically different distributions of true positives, false positives, and disputes across implementations, revealing where specification--implementation alignment diverges.

\subsection{Threat Model}
\label{sec:design:threat-model}

\noindent\textbf{Distributed-systems threat surface.}
\sys targets multi-implementation protocols whose safety, liveness, and integrity emerge only when every client honours the specification consistently. A specification-divergence bug in any one client creates a network-scoped attack surface: an adversary delivering crafted inputs (blocks, attestations, peer messages, RPC, engine-API responses) to a single client can fork the network, halt finality, corrupt cross-client state, or distort reward/slashing accounting. The blast radius is the network, not the host.

\noindent\textbf{Implementation-level scope.}
\sys targets implementation vulnerabilities (correctness failures, denial of service, information disclosure, privilege escalation) in specification-governed software.
Specifications and the bug-bounty scope document are \emph{trusted}; inter-component channels are \emph{semi-trusted}; external inputs are \emph{untrusted} (Table~\ref{tab:trust-model}).
\sys does not target specification-level flaws, hardware attacks, or vulnerabilities that require dynamic execution to observe (e.g., race conditions). We assume specifications in machine-readable form with RFC~2119 modal verbs, accessible source code, and a bug-bounty scope document.

\begin{table}[t]
\centering
\caption{Trust boundary model used throughout the audit pipeline. Each boundary is mapped to a trust level that Phase~6's trust-boundary gate checks against attack-path sources.}
\label{tab:trust-model}
\footnotesize
\setlength{\tabcolsep}{4pt}
\begin{tabular}{lll}
\toprule
\textbf{Boundary} & \textbf{Trust Level} & \textbf{Rationale} \\
\midrule
Specification documents & TRUSTED & Correctness oracle \\
Bug bounty scope & TRUSTED & Defines audit perimeter \\
Internal component APIs & SEMI\_TRUSTED & Malformed but valid data \\
External/user-facing inputs & UNTRUSTED & Attacker-controlled \\
\bottomrule
\end{tabular}
\end{table}

\subsection{Property Vocabulary}
\label{sec:design:vocab}

Phase~3 synthesizes each property from six sources---STRIDE threats, trust-boundary
obligations, specification assumptions, subgraph invariants (\texttt{INV-*}), state
transitions, and optimization-correctness constraints---and assigns one of four types.
Table~\ref{tab:property-types} summarizes the vocabulary.

\begin{table}[t]
\centering
\caption{The four property types in \sys's vocabulary.}
\label{tab:property-types}
\small
\begin{tabular}{l>{\raggedright\arraybackslash}p{2.2cm}>{\raggedright\arraybackslash}p{3.0cm}}
\toprule
\textbf{Type} & \textbf{Definition} & \textbf{Example} \\
\midrule
Invariant & Holds at all times & Nodes maintain $\geq$ \texttt{CUSTODY\_REQUIREMENT} columns \\
\addlinespace
Precondition & Holds before an operation & Input length equals commitment count before verification \\
\addlinespace
Postcondition & Holds after operation completes & New custody set covers required subnets after rotation \\
\addlinespace
Trust assumption & Assumed but not enforced by component (e.g., a downstream trust boundary) & Execution layer returns valid block data via Engine API \\
\bottomrule
\end{tabular}
\end{table}

\subsection{Property-Grounded Audit (Phase 5)}
\label{sec:design:phase5}

Phase~5 is the core reasoning phase: for each property, the auditor attempts to construct a proof that the property holds; gaps constitute findings.
We characterize this as \emph{Hoare-style proof-attempt reasoning}---the LLM treats every property as an assertion to be discharged at the relevant program points, and emits a finding only when no discharging argument is reachable.
The pre/post/invariant/assumption typing maps directly onto this discharge style (preconditions at call entry, postconditions at return, invariants over the function body, assumptions at trust-boundary interfaces).
Figure~\ref{fig:phase5} shows the audit flow; Figure~\ref{fig:audit-pseudocode} the detailed procedure.

\begin{figure}[t]
\small
\begin{algorithmic}[1]
\Require Property $P$ with assertion $A$, code scope $C$
\Ensure Finding $F$ or verification \textsc{Pass}
\Procedure{Audit}{$P$, $C$}
  \State $\mathit{sub\_claims} \gets \Call{Decompose}{A}$
  \State $\mathit{code\_map} \gets \Call{Map}{\mathit{sub\_claims}, C}$ \Comment{Sub-phase 1}
  \For{each claim $s$ in $\mathit{sub\_claims}$}
    \State $\mathit{proof} \gets \Call{Prove}{s, \mathit{code\_map}[s]}$ \Comment{Sub-phase 2}
    \If{$\mathit{proof}$.fails}
      \State $\mathit{gap} \gets \mathit{proof}$.failure\_point
      \State $\mathit{attack} \gets \Call{StressTest}{\mathit{gap}, C}$ \Comment{Sub-phase 3}
      \If{$\mathit{attack}$.plausible}
        \State \Return \Call{Finding}{$P$.id, \textsc{Vuln}, $\mathit{gap}$, $\mathit{attack}$}
      \Else
        \State \Return \Call{Finding}{$P$.id, \textsc{Potential}, $\mathit{gap}$, $\emptyset$}
      \EndIf
    \EndIf
  \EndFor
  \State \Return \textsc{Pass}
\EndProcedure
\end{algorithmic}
\caption{Property-grounded audit procedure for a single property.}
\label{fig:audit-pseudocode}
\end{figure}

\begin{figure*}[t]
\centering
\includegraphics[width=0.85\textwidth]{phase5.png}
\caption{Phase~5 audit flow for a single property. Each property undergoes three sub-phases: \emph{Map} (decompose assertion into sub-claims), \emph{Prove} (verify each sub-claim against code), and \emph{Stress-Test} (validate or challenge the conclusion).}
\label{fig:phase5}
\end{figure*}

Each property is processed in three sub-phases.
\emph{Map}: decompose the assertion into verifiable sub-claims, read the enforcement code (full function bodies plus callers/callees), and link each sub-claim to its responsible code.
\emph{Prove}: verify input coverage, path coverage, concurrency safety, temporal validity, and implementation-pattern obligations (e.g., cache keys and deduplication keys must be computed from complete inputs); gaps are recorded as findings.
\emph{Stress-test}: re-examine assumptions if the proof succeeded, or construct a concrete attack path if it failed; findings without a plausible path are downgraded to \texttt{potential-vulnerability}.
Phase~5 is LLM-driven evidence construction with structured reasoning, not formal verification.
The proof-attempt framing was chosen over an adversarial ``find bugs'' framing after preliminary experiments showed the latter produced an 88\% false-positive rate: without a structured claim to disprove, the model produced numerous speculative findings with weak grounding, whereas the proof-attempt framing forces it to commit to a verifiable claim (the property's assertion) before reporting any gap, eliminating speculative reasoning at the source.

\begin{sloppypar}
\noindent\textbf{Worked example.}
For the property ``The KZG batch verification challenge must be computed from the deduplicated commitment array,'' Map identifies \texttt{compute\_verify\_cell}\allowbreak\texttt{\_kzg\_proof}\allowbreak\texttt{\_batch\_challenge} at line~877; Prove discovers that the function receives \texttt{commitments\_bytes} (with duplicates) rather than \texttt{unique\_commitments}, breaking the verification scheme; Stress-Test confirms a crafted duplicate set causes the challenge hash to diverge from the specification, enabling proofs to pass incorrectly. The result is a High-severity cryptographic correctness bug missed by 366 human auditors in the concurrent contest (\S\ref{sec:eval:rq1}).
\end{sloppypar}

\subsection{Severity-Preserving Review (Phase 6)}
\label{sec:design:phase6}

Security auditing faces an asymmetric risk: a missed H/M/L true positive is costlier than a false positive requiring triage time. \sys's review phase therefore prioritizes \emph{severity preservation}: it may only dispute findings through three narrow, mechanical gates designed to be recall-safe for H/M/L findings.

\noindent\textbf{Gate~1: Dead Code.} If a flagged function has zero non-test callers (and is not part of a public API), the finding is classified \texttt{DISPUTED\_FP}.
\noindent\textbf{Gate~2: Trust Boundary.} If the attack path's data source is \texttt{TRUSTED} or \texttt{SEMI\_TRUSTED} and no alternative untrusted input path exists, classified \texttt{DISPUTED\_FP}.
\noindent\textbf{Gate~3: Scope Check.} Findings targeting scope-excluded categories, pre-existing issues, or out-of-scope components are classified \texttt{DISPUTED\_FP}.

Only these three gates may produce a \texttt{DISPUTED\_FP} verdict; no design-intent or code-quality reasoning may dispute a finding. The three-gate design was simplified from a five-gate prototype after two gates (Spec Cross-Reference, Exploitability) caused recall loss on informational true positives at negligible precision benefit (\S\ref{sec:eval:rq1}). Surviving findings receive severity calibration against the scope document's impact thresholds.

\section{Evaluation}
\label{sec:evaluation}

\subsection{Research Questions}
\label{sec:eval:rq}

\noindent\textbf{RQ1: How effective is \sys at auditing specification-governed implementations, and what explains its successes and failures?} We measure detection effectiveness (recall, precision) and require that both successes and failures be traceable to specific pipeline mechanisms.

\noindent\textbf{RQ2: How does \sys compare to code-driven auditing, and does the property-centered representation produce qualitatively different outcomes?} We test for (i)~discovery of bugs beyond the established ground truth and (ii)~qualitatively different security judgments on disputed patterns.

\subsection{Experimental Setup}
\label{sec:eval:setup}

\noindent\textbf{RQ1 benchmark (Sherlock).}
We evaluate on the Sherlock Ethereum Fusaka Audit Contest, targeting 10 Ethereum implementations of EIP-7594 (PeerDAS) and EIP-7691; Table~\ref{tab:sherlock-bench} summarizes the benchmark. The contest received 366 submissions from professional auditors, of which 15 were judged valid at H/M/L severity (5~H, 2~M, 8~L). The 10 targets span six languages (Go, Rust, Nim, TypeScript, C\#, C) and include 5 consensus-layer clients (Lighthouse, Prysm, Grandine, Nimbus, Lodestar), 2 execution clients (Nethermind, Reth), 1 EVM library (Alloy EVM), and 2 cryptographic libraries (c-kzg-4844, rust-eth-kzg).

\begin{table}[t]
\centering
\caption{Sherlock Ethereum Fusaka Audit Contest benchmark characteristics.}
\label{tab:sherlock-bench}
\footnotesize
\setlength{\tabcolsep}{4pt}
\begin{tabular}{ll}
\toprule
\textbf{Characteristic} & \textbf{Value} \\
\midrule
Target implementations & 10 (5 consensus + 2 execution clients + 3 libraries) \\
Programming languages & Go, Rust, Nim, TypeScript, C\#, C \\
Contest submissions & 366 \\
Valid issues (H/M/L) & 15 (5 High, 2 Medium, 8 Low) \\
Severity thresholds & Network-impact-based \\
\bottomrule
\end{tabular}
\end{table}
The ground truth is established by independent expert adjudication, not synthetic bug injection.
An earlier version of \sys was deployed live in this contest (Sep--Oct~2025) and placed 6th overall with the largest valid-issue count (17) among participating teams; the analyses below are from a controlled post-contest re-experiment, separating methodology from operator effects.
Direct comparison with individual contest participants is precluded by confidentiality; we therefore evaluate absolutely against the adjudicated ground truth and reserve head-to-head baseline comparison for RQ2.

\noindent\textbf{RQ2 benchmark (RepoAudit).}
We evaluate on the RepoAudit benchmark~\cite{repoaudit}: 15 open-source C/C++ projects
with 35 non-disputed ground-truth bugs (null-pointer dereferences, memory leaks, and
use-after-free) confirmed by developer fixes, plus 5 disputed bugs.
We compare against published baseline numbers from Guo et al.~\cite{repoaudit} (four
RepoAudit model configurations) and two industrial static analyzers (Meta Infer, Amazon
CodeGuru).
We do not report recall comparatively because the ground truth was constructed from
RepoAudit's own discoveries, making recall comparison structurally unfair.

\noindent\textbf{Metrics.}
We report precision (non-FP rate), recall on H/M/L contest issues, and a strict H/M/L-match rate.
Issue matching uses LLM-assisted semantic matching followed by two-author manual verification (initial agreement 93\%, consensus on disagreements).
A match is accepted only when both root cause and security impact align; partial matches (same code location but different impact) are excluded from recall. Many-to-one matches count the issue recovered if any finding matches, with each finding counted independently for precision.
Findings are labeled true positive (\texttt{tp}, \texttt{tp\_info}, \texttt{fixed}, \texttt{partially\_fixed}, \texttt{potential-info}) or false positive (\texttt{fp\_invalid}, \texttt{fp\_review}); Appendix~\ref{app:eval:setup} provides the full label definitions.
We report precision at three granularities (strict H/M/L match, confirmed-useful, broad non-FP) to expose effectiveness under different TP interpretations.

\noindent\textbf{Caveats.}
Three methodological choices condition our numbers and warrant explicit acknowledgement: (i)~issue matching uses LLM-assisted semantic matching with two-author manual verification (initial agreement 93\%), introducing bounded but non-zero matching uncertainty; (ii)~the Sherlock severity adjudication is contest-organizer-defined and conservatively requires network-wide impact for Medium/High, so operationally significant local bugs surface as Informational and depress strict precision; and (iii)~the RepoAudit ground truth was constructed from RepoAudit's own discoveries, so we report beyond-GT candidates rather than recall against it. Appendix~\ref{app:discussion} elaborates on each.

\noindent\textbf{Implementation.}
\sys is a Python systems framework ($\sim$1{,}900 lines of async orchestration plus 1{,}373 lines of analysis prompts across 9 workers); workers run in parallel via \texttt{asyncio} with circuit-breaker abort and per-batch checkpointing.
Phases~1--3 use Claude Opus; Phases~4--6 use Claude Sonnet; Phase~4 uses Tree-sitter via MCP for symbol resolution.

\noindent\textbf{Cost and wall-clock.}
Total API cost was \$400--620 for RQ1 ($\sim$\$27--41 per recovered H/M/L bug, or \$21--33 per unique bug including the 4 fix-confirmed novel bugs) and \$81.05 (\$1.69/bug) for the RQ2 primary configuration (Sonnet~4.5); multi-model sensitivity (Sonnet~4, DeepSeek~R1) cost \$93--101 per model.
Phase~5 finishes in $\sim$42~min wall-clock for all 10 RQ1 targets under parallel execution (4 workers/target, 64 concurrent items).
Evaluation scripts and all generated artifacts are released as supplementary material (Open Science, p.\,\pageref{sec:openscience}).

\subsection{RQ1: Effectiveness and Structured Failure Analysis}
\label{sec:eval:rq1}

\subsubsection{Detection Effectiveness}

Phase~5 produces 647 audit outputs; 545 are classified as not-a-vulnerability or
out-of-scope, leaving 102 positive findings.
Phase~6 filters 30 as \texttt{DISPUTED\_FP}, yielding 72 post-review findings.
Table~\ref{tab:rq1-headline} presents the headline detection results;
Table~\ref{tab:rq1-accounting} (Appendix~\ref{app:eval:accounting}) provides a
full accounting reconciliation.

\begin{table}[t]
\centering
\caption{\sys detection effectiveness on the Sherlock benchmark. The gap between strict precision (26.4\%) and broad precision (66.7\%) is explained by the property neighborhood effect (one issue detected by multiple complementary properties) and by informational-severity/independently-fixed findings outside the contest's severity-gated scope.}
\label{tab:rq1-headline}
\small
\begin{tabular}{lr}
\toprule
\textbf{Metric} & \textbf{Value} \\
\midrule
Total findings (Phase 5, pre-review) & 102 \\
Total findings (Phase 6, post-review) & 72 \\
Confirmed FPs (post-review) & 24 (33.3\%) \\
H/M/L recovered (expert-augmented) & 15 / 15 (100\%) \\
H/M/L recovered (automated-only) & 8 / 15 (53\%) \\
Novel bugs confirmed by fix commits & 4 \\
\midrule
\multicolumn{2}{l}{\emph{Precision at three granularities (post-review, N=72):}} \\
\quad Strict (H/M/L match) & 26.4\% (19/72) \\
\quad Confirmed-useful & 59.7\% (43/72) \\
\quad Broad (non-FP rate) & 66.7\% (48/72) \\
\bottomrule
\end{tabular}
\end{table}

In the expert-augmented configuration, \sys recovers all 15 in-scope H/M/L vulnerabilities (5~High, 2~Medium, 8~Low). Phase~6 improves broad precision from 56.9\% to 66.7\% while preserving 100\% recall on H/M/L (raising F1 from 72.5\% to 80.0\%); the 10 TPs filtered by Phase~6 are all informational/partial (8~\texttt{tp\_info}, 1~\texttt{partially\_fixed}, 1~\texttt{potential-info}), confirming the severity-preserving design.

\begin{figure}[t]
\centering
\includegraphics[width=\columnwidth]{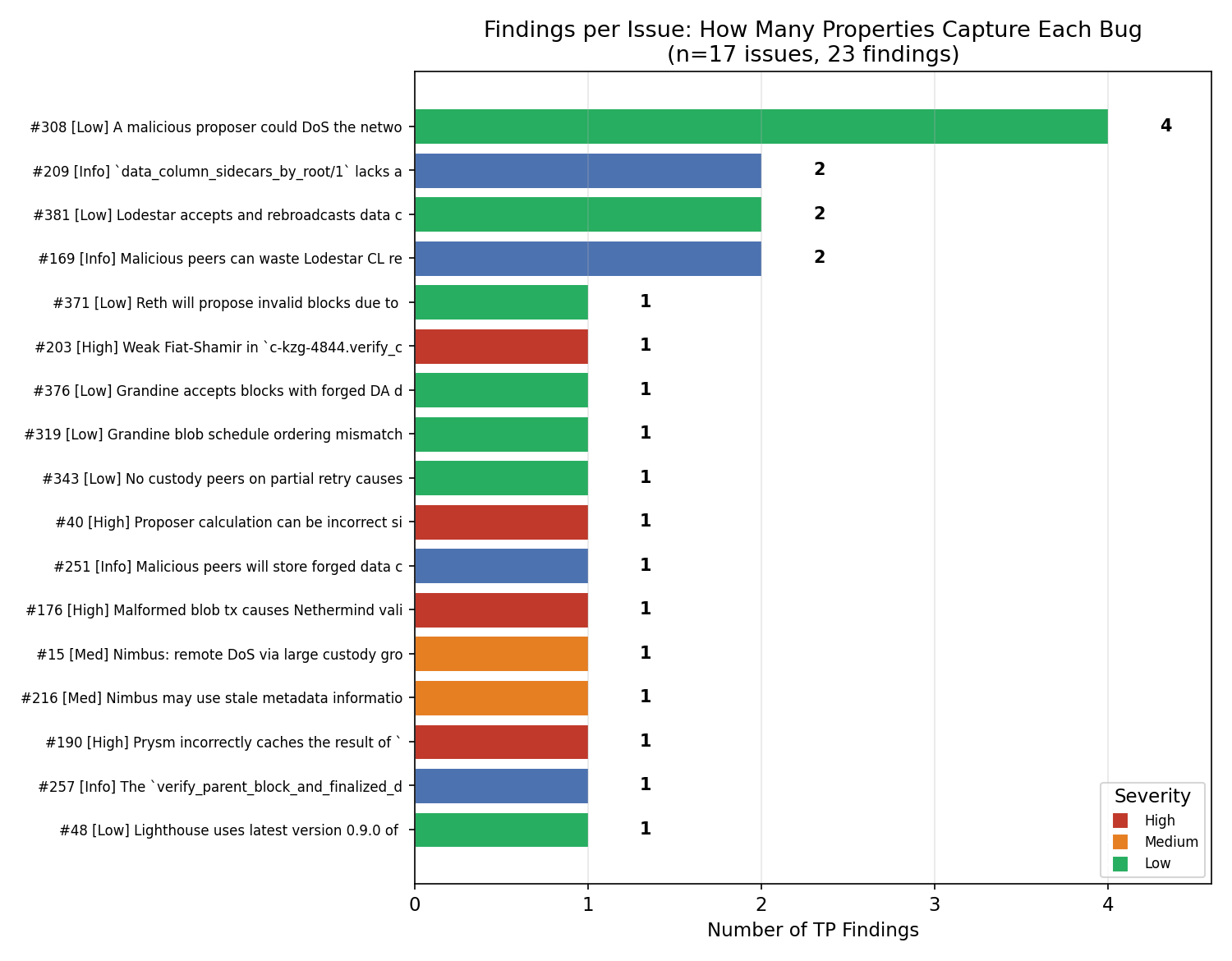}
\Description{Findings per issue.}
\caption{Findings per issue: the property neighborhood effect. Eleven of 15 matched issues are covered by two or more complementary properties, explaining both the high recall and the gap between finding-level and cluster-level precision.}
\label{fig:rq1-fpi}
\end{figure}

\noindent\textbf{Property neighborhoods explain the strict/broad gap.}
The strict H/M/L match rate (26.4\%) reflects a \emph{property neighborhood effect}: 11 issues are matched by 2+ properties (e.g., \#308 and \#331 each by 4; \#169 by 3); Figure~\ref{fig:rq1-fpi} visualizes the full per-issue distribution. Grouping findings by issue (Table~\ref{tab:rq1-cluster}) yields cluster-level strict precision of 48.7\% (15 issues carried by 19 clusters rather than 72 alerts), confirming genuine redundancy rather than alert duplication---and this many-to-one structure also explains the high recall: the probability that at least one neighborhood property matches the root cause is high.

\begin{table}[t]
\centering
\caption{Finding-level vs cluster-level evaluation. Cluster-level strict precision (48.7\%) nearly doubles the finding-level rate (26.4\%), confirming that the property neighborhood provides genuine redundancy rather than alert duplication.}
\label{tab:rq1-cluster}
\footnotesize
\setlength{\tabcolsep}{4pt}
\begin{tabular}{lll}
\toprule
\textbf{Level} & \textbf{Precision (broad / strict)} & \textbf{Recall} \\
\midrule
Finding-level (N=72) & 66.7\% / 26.4\% & 15/15 (100\%) \\
Cluster-level (N=39) & --- / 48.7\% & 15/15 (100\%) \\
\bottomrule
\end{tabular}
\end{table}

\begin{figure}[t]
\centering
\includegraphics[width=\columnwidth]{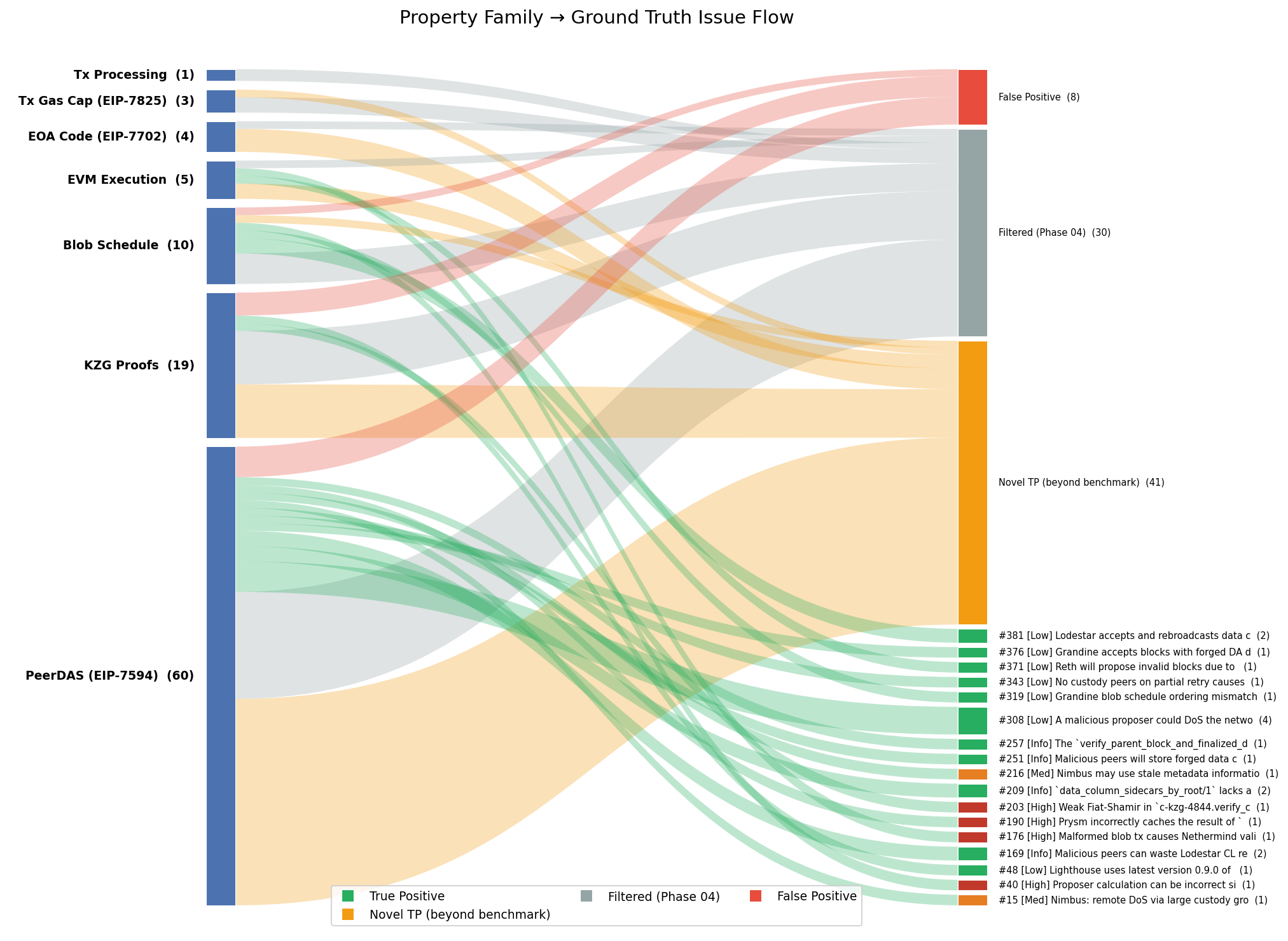}
\Description{Sankey diagram from property families to ground-truth issues.}
\caption{Sankey flow from the property families that recover the 15 ground-truth issues, across the 10 Ethereum audit targets. Phase 3 generates 7 property families in total; 5 are reused across two or more implementations, and 4 distinct families together account for all 15 ground-truth recoveries. The horizontal density around a small subset of issues visualizes the \emph{property neighborhood effect}: complementary properties jointly characterize each high-impact issue, so cluster-level precision (48.7\%) nearly doubles the finding-level rate---an effect that is visible only when properties are held constant across implementations.}
\label{fig:rq1-sankey}
\end{figure}

\sys also independently discovered 4 bugs confirmed by developer fix commits, none of which surfaced in the 366 adjudicated contest submissions:
(A)~the c-kzg-4844 KZG batch-verification challenge computed from the original (not deduplicated) commitment array (fix: \texttt{f18ba082});
(B)~an inverted logic error plus missing validation in Lodestar (\texttt{3b98c59c});
(C)~an unchecked array access in Nimbus reachable from RPC/Engine API (\texttt{b3a3f3f9});
(D)~a wrong subnet parameter computation plus missing cell-count validation in Prysm (\texttt{b5bdd65f}).
All four share a structural property: each violates a specification-derived expectation that the implementation's local code does not encode---which is why a code-driven pipeline lacks the vocabulary to surface them. Bug~A is the most striking: a cryptographic correctness violation in a library shared by multiple Ethereum clients, where local code passes type and length checks but breaks the polynomial commitment scheme's mathematical premise. Bug~B is a cross-field constraint between two locally-correct fields whose required equality is defined only by the spec. Bug~C is reachable from an external trust boundary (RPC/Engine API) that local code reads as trusted, missing a spec-defined input-validation obligation. Bug~D is a protocol-parameter divergence at fork epochs visible only when the canonical schedule is held as oracle. The comparison is not fully controlled---the post-contest configuration had structured specification access and substantial compute budget, while contest participants operated under time pressure.

\subsubsection{Cross-Implementation Patterns}
\label{sec:eval:rq1-crossimpl}

\begin{figure}[t]
\centering
\includegraphics[width=\columnwidth]{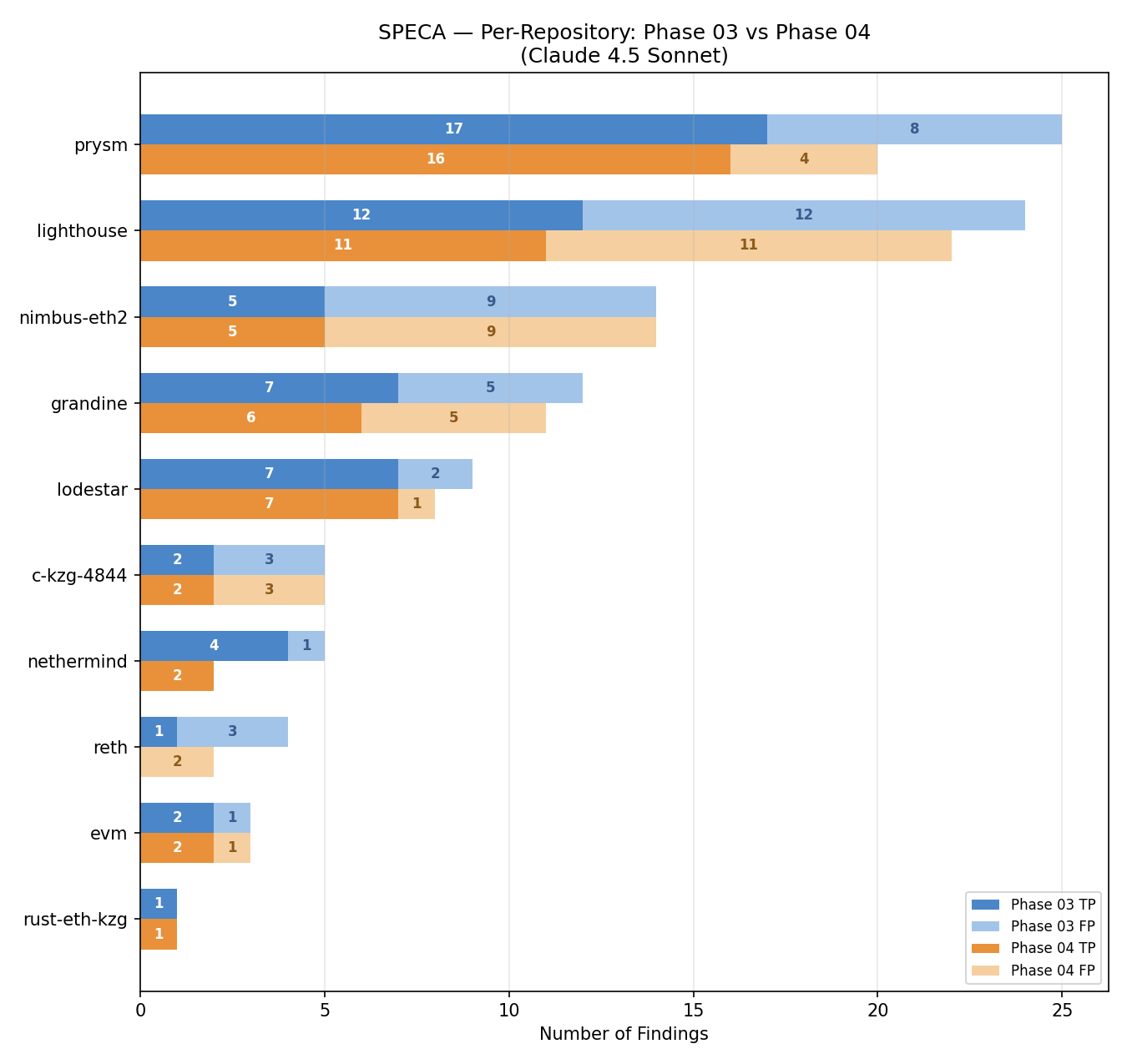}
\Description{Per-repository finding counts.}
\caption{Per-repository finding distribution (Phase~5 vs Phase~6). The same property vocabulary produces visibly different TP/FP profiles across the 10 audit targets---a direct consequence of holding security expectations constant while varying the implementation. Code-local tools cannot produce this view.}
\label{fig:rq1-perrepo}
\end{figure}

Because \sys applies a shared property vocabulary across all 10 implementations, we can analyze where the same expectations produce different outcomes. Figure~\ref{fig:rq1-sankey} visualizes how the 7 property families map to ground-truth issues across all targets: \emph{5 of the 7 families fire on 2 or more implementations}, which is the operational evidence for the shared-vocabulary contribution---without it, no such reuse measurement would be definable. Eleven issues recur across multiple repositories (e.g., \#308 in Grandine, Lighthouse, Nimbus, Prysm; \#169 in Grandine, Lodestar, Prysm). As a case study, \texttt{PROP-6a4369e9-inv-014} asserts that blob parameters at fork boundaries must be computed from the canonical schedule rather than local config defaults: Grandine, Lighthouse, and Prysm use a divergent local default (all three TP, matched to contest issue~\#11), while Nimbus correctly reads the canonical schedule (PASS). A single property thus functions simultaneously as a true-positive detector in deviating implementations and as a conformance certificate in compliant ones---a controlled comparison code-local tools cannot achieve. Figure~\ref{fig:rq1-perrepo} shows the per-repository finding distribution under the shared vocabulary; Appendix~\ref{app:eval:crossimpl} provides further breakdowns.

\subsubsection{Automated-Only vs.\ Expert-Augmented Configurations}
\label{sec:eval:rq1-config}

Of the 15 detecting properties, 8 were auto-generated by Phases~1--3 and 7 required manual creation, clustering in cryptographic invariants (KZG edge cases, BLS12-381 identity element) and protocol lifecycle rules (custody bounds, cache key completeness, fork-transition refresh)---areas requiring mathematical domain knowledge or multi-specification cross-referencing.
The automated-only configuration thus recovers 8/15 (53\%); the expert-augmented configuration recovers 15/15 (100\%) (Table~\ref{tab:configs}).

\begin{table}[t]
\centering
\caption{Automated-only vs.\ expert-augmented configurations. The 7 manual properties amortize across all 10 implementations, turning expert knowledge injection into a one-time cost per specification corpus.}
\label{tab:configs}
\small
\begin{tabular}{l>{\raggedright\arraybackslash}p{2.4cm}rr}
\toprule
\textbf{Config} & \textbf{Properties} & \textbf{H/M/L} & \textbf{Cov.} \\
\midrule
Automated-only & Auto (Phases 1--3) & 8 / 15 & 53\% \\
Expert-augmented & Auto + 7 manual & 15 / 15 & 100\% \\
\bottomrule
\end{tabular}
\end{table}

The current evidence supports characterizing \sys as a \emph{specification-anchored audit framework with expert augmentation}. The framework's value lies in the property-centered representation, the structured proof-attempt audit pipeline, and the severity-preserving review filter---each fully automated and property-source-agnostic. A manually authored property passes through the same pipeline as an auto-generated one, and the 7 manual properties amortize across all 10 implementations: expert knowledge injection has high leverage in multi-implementation settings. Improving automated extraction for cryptographic and protocol-lifecycle properties is a concrete research frontier (Appendix~\ref{app:eval:spec}).

\subsubsection{Structured False-Positive Analysis}
\label{sec:eval:rq1-fp}

\sys produces 44 total FPs across both phases (24 surviving Phase~6, 20 correctly filtered). We selected 16 FPs for detailed pipeline tracing; all trace to three identifiable pipeline limitations (Table~\ref{tab:fp-taxonomy}).

\begin{table}[t]
\centering
\caption{False-positive root causes with phase origin (N=16 deeply analyzed from a population of 44 total FPs). Each root cause maps to a specific pipeline phase and to a concrete, implementable improvement.}
\label{tab:fp-taxonomy}
\footnotesize
\setlength{\tabcolsep}{3pt}
\begin{tabular}{l>{\raggedright\arraybackslash}p{2.1cm}rr}
\toprule
\textbf{Root Cause} & \textbf{Phase Origin} & \textbf{N} & \textbf{\%} \\
\midrule
Trust boundary misunderstanding & Phase 3 & 8 & 50.0 \\
Code reading error & Phase 5 & 6 & 37.5 \\
Specification misinterpretation & Phase 3 & 2 & 12.5 \\
\midrule
\textbf{Total} & & \textbf{16} & \textbf{100} \\
\bottomrule
\end{tabular}
\end{table}

\emph{Trust-boundary misunderstanding} (50\%, Phase~3): trusted inter-component channels (e.g., consensus--execution JWT) treated as untrusted. \emph{Code-reading error} (37.5\%, Phase~5): misreading implementation logic (e.g., confusing a bounds check with a missing one). \emph{Specification misinterpretation} (12.5\%, Phase~3): properties from ambiguous spec text the protocol team reads differently. Each maps to a concrete improvement; the Phase~6 gate filters half the trust-boundary FPs and two-thirds of code-reading FPs but none of the misinterpretation ones. Appendix~\ref{app:eval:fp} provides per-category traces and gate breakdowns.

\noindent\textbf{Property-type ablation.}
Ablating precision by property type over an N=88 subset (Table~\ref{tab:prop-ablation}): invariants drive 76\% of findings at 75\% precision; preconditions reach 100\% (small sample); postconditions are moderate (50\%); assumptions are too noisy for reliable auditing (0\%) and are best treated as an exploratory mode. The mechanism behind the assumption result is structural: a trust assumption asserts a property the audited component does \emph{not} itself enforce (e.g., ``Engine API returns valid block data''), so Phase~5 has no in-scope discharge target and any flagged deviation is by design out of the audited code's responsibility---most assumption-flagged cases on inspection turn out to be invariants miscategorized at Phase~3. Excluding assumption properties leaves precision essentially unchanged (67.1\%) and recall stable (27.1\% strict), confirming that assumptions contribute negligibly to actionable detections. Postcondition and assumption generation are the concrete research frontiers for the automated-only configuration.

\begin{table}[t]
\centering
\caption{Property-type ablation (N=88).}
\label{tab:prop-ablation}
\small
\begin{tabular}{lrrrr}
\toprule
\textbf{Property Type} & \textbf{N} & \textbf{TP} & \textbf{FP} & \textbf{Precision} \\
\midrule
Invariant & 67 & 18 & 6 & 75.0\% \\
Precondition & 11 & 4 & 0 & 100.0\% \\
Postcondition & 5 & 1 & 1 & 50.0\% \\
Assumption & 5 & 0 & 1 & 0.0\% \\
\bottomrule
\end{tabular}
\end{table}

\noindent\textbf{Phase~6 verdict breakdown.}
Table~\ref{tab:phase6-verdicts} reports the Phase~6 verdict distribution and retrospective TP rate over the 102 pre-review findings. The \texttt{DISPUTED\_FP} verdict filters two-thirds of the false positives it flags while preserving H/M/L recall---the severity-preserving design point. The three-gate filter was simplified from a five-gate prototype after the dropped gates (Spec Cross-Reference, Exploitability) filtered informational TPs at zero precision. Figure~\ref{fig:rq1-gates} (appendix) visualizes per-gate effectiveness.

\begin{table}[t]
\centering
\caption{Phase~6 verdict distribution and retrospective TP rate (N=102). \texttt{DISPUTED\_FP}'s 33.3\% TP rate means the filter achieves 66.7\% precision as a false-positive detector while incurring only informational recall loss.}
\label{tab:phase6-verdicts}
\small
\begin{tabular}{lrrr}
\toprule
\textbf{Verdict} & \textbf{Count} & \textbf{TP} & \textbf{TP Rate} \\
\midrule
CONFIRMED\_VULNERABILITY & 39 & 29 & 74.4\% \\
CONFIRMED\_POTENTIAL & 24 & 13 & 54.2\% \\
DOWNGRADED & 8 & 5 & 62.5\% \\
DISPUTED\_FP & 30 & 10 & 33.3\% \\
NEEDS\_MANUAL\_REVIEW & 1 & 1 & 100.0\% \\
\bottomrule
\end{tabular}
\end{table}

\subsubsection{Specification Contribution}
\label{sec:eval:rq1-spec}

To test whether \sys's detections truly depend on specification-derived information, we retrospectively classified each of the 15 detecting properties as \emph{spec-essential} or \emph{code-inferrable} and traced its provenance to a specification source. All 15 were judged spec-essential, falling into four recurring patterns (Table~\ref{tab:spec-essential}).

\begin{table}[t]
\centering
\caption{Retrospective spec-essentiality classification of all 15 matched issues. All 15 were judged spec-essential; 0 were judged code-inferrable. Authors' assessment.}
\label{tab:spec-essential}
\footnotesize
\setlength{\tabcolsep}{4pt}
\begin{tabular}{l>{\raggedright\arraybackslash}p{3.4cm}r}
\toprule
\textbf{Category} & \textbf{Issues} & \textbf{Count} \\
\midrule
Spec-defined algorithm & \#40, \#176, \#210, \#308, \#319 & 5 \\
Mathematical invariant & \#48, \#203 & 2 \\
Cross-field constraint & \#190, \#343, \#371, \#376, \#381 & 5 \\
Protocol parameter & \#15, \#109, \#216 & 3 \\
\bottomrule
\end{tabular}
\end{table}

The categories illustrate why code-local analysis cannot synthesize these properties. Issue~\#40 recomputes the proposer index correctly by the \emph{original} algorithm but the Fulu spec redefines the function to use a cached lookahead array; the bug is non-conformance with the new spec. Issue~\#48 involves a KZG commitment equal to the BLS12-381 identity element, which passes group-membership checks but violates a mathematical invariant only the specification defines. Issue~\#376 involves \texttt{prev\_randao} and \texttt{randao\_mix}, individually correct but required to be equal by the spec's \texttt{process\_execution\_payload}---a cross-field constraint no single function reveals.
The remaining categories follow the same pattern at scale: the four other cross-field constraints (\#190, \#343, \#371, \#381) all require correlating fields whose individual local invariants hold; the three protocol-parameter issues (\#15, \#109, \#216) involve constants whose authoritative value lives in the specification's parameters table and that the implementation hard-codes inconsistently; and the four other spec-defined-algorithm issues (\#176, \#210, \#308, \#319) all instantiate the same ``algorithm redefined by spec but not by the implementation'' pattern as \#40.
All 15 trace through the property$\rightarrow$subgraph$\rightarrow$spec-section provenance chain to specification sources; 2 of 15 terminate at a specific \texttt{INV} label (e.g., issue~\#319 traces through subgraph \texttt{SG-001} for \texttt{get\_blob\_parameters} to \texttt{INV-004}: ``if no schedule entry matches, defaults to ELECTRA parameters''), making the spec-essential claim mechanically auditable rather than retrospective assertion. This supports characterizing \sys's gains as representation-induced rather than model-capability-induced.

\subsection{RQ2: Comparison Against Code-Driven Baselines}
\label{sec:eval:rq2}

We run \sys on the 15 C/C++ projects from RepoAudit using the same pipeline as RQ1. Table~\ref{tab:rq2-main} compares against published baselines, organized into a same-backbone comparison (DeepSeek~R1), \sys's best configuration, and other configurations.

\begin{table*}[t]
\centering
\caption{Comparison on the RepoAudit benchmark. TP numbers for RepoAudit are published by Guo et al.\ and include both GT-matched bugs and additional discoveries. For \sys, we omit an aggregate TP count because computing it requires estimating GT matches---an inherently approximate procedure. Recall is not reported because the GT was constructed from RepoAudit's own discoveries. $^b$New candidates = author-validated bugs beyond the established GT (Level~D validation, see Appendix~\ref{app:eval:rq2}).}
\label{tab:rq2-main}
\small
\begin{tabular}{llrrrrr}
\toprule
& \textbf{Method} & \textbf{TP} & \textbf{FP} & \textbf{Precision} & \textbf{New cand.$^b$} & \textbf{Cost} \\
\midrule
\multicolumn{7}{l}{\emph{Partially controlled comparison (DeepSeek R1)}} \\
& RepoAudit (DeepSeek R1) & 41 & 6 & 87.2\% & (in TP) & \$8.55 \\
& \sys (DeepSeek R1) & --- & 15 & 72.7\% & 7 & \$93.51 \\
\midrule
\multicolumn{7}{l}{\emph{Latest models}} \\
& RepoAudit (Claude 3.7 Sonnet) & 40 & 5 & 88.9\% & (in TP) & \$23.85 \\
& \textbf{\sys (Sonnet 4.5)} & \textbf{---} & \textbf{6} & \textbf{88.9\%} & \textbf{12} & \$81.05 \\
\midrule
\multicolumn{7}{l}{\emph{Other configurations}} \\
& Amazon CodeGuru & 0 & 18 & 0.0\% & 0 & --- \\
& Meta Infer & 7 & 2 & 77.8\% & 0 & free \\
& RepoAudit (o3-mini) & 36 & 9 & 80.0\% & (in TP) & \$4.50 \\
& RepoAudit (Claude 3.5 Sonnet) & 40 & 11 & 78.4\% & (in TP) & \$38.10 \\
& \sys (Sonnet 4) & --- & 13 & 81.2\% & 18 & \$100.68 \\
\bottomrule
\end{tabular}
\end{table*}

\noindent\textbf{Matched-backbone comparison.}
Under a partially controlled same-backbone comparison (DeepSeek~R1; agent architecture and retrieval still differ), \sys reports 7 author-validated candidate bugs beyond ground truth while RepoAudit achieves higher precision (87.2\% vs.\ 72.7\%)---consistent with a representation effect but not isolating it. With Sonnet~4.5, \sys's precision improves to 88.9\% (matching the best RepoAudit config) while discovering 12 author-validated bugs beyond ground truth; Sonnet~4, intermediate in capability, discovers \emph{more} candidates (18) at 81.2\% precision---an apparent inversion explained by the \emph{property-adherence effect} elaborated below: stronger models stay within the stated property scope, while weaker models drift into adjacent code patterns where some genuine bugs happen to live. Combined with 100\% recall on the 35 non-disputed ground-truth bugs, Sonnet~4.5 reaches $F_1{=}0.94$, on par with RepoAudit's strongest published configuration (Claude~3.7 Sonnet, $F_1{=}0.94$); the differentiation is qualitative, in the form of beyond-GT discoveries and disputed-bug verdicts.

\noindent\textbf{Divergent security judgments.}
\sys and RepoAudit produce qualitatively different verdicts on 5 disputed patterns, reflecting a methodological difference: \sys applies an exploitability-centered interpretation (vulnerability only with a concrete attack path), while RepoAudit flags best-practice violations regardless of exploitability. For example, \texttt{RA-M2-O1} (memcached \texttt{server\_sockets} leak) is not exploitable---both call sites immediately invoke \texttt{exit()}---so \sys classifies it as not-a-vulnerability while RepoAudit flags it as a leak. We report these as divergent security judgments rather than declaring either tool correct.
This judgment is model-dependent: DeepSeek~R1 rejects all 5, Sonnet~4.5 rejects 4 of 5, Sonnet~4 detects 4 of 5---consistent with the principle that exploitability-centered classification requires counterfactual reasoning where larger models improve most. The takeaway is that a specification-grounded pipeline surfaces methodological disagreement explicitly as a property-level verdict, available for human adjudication.

\noindent\textbf{Beyond-GT candidates: discovery rate and diversity.}
Across all model configurations \sys discovers 7--18 author-validated candidate bugs beyond ground truth. Sonnet~4 contributes the highest count (18, listed in Table~\ref{tab:new-bugs}, Appendix~\ref{app:eval:rq2}), spanning three bug types (use-after-free 9, memory leak 6, NPD 3) and eleven projects (with the Linux kernel contributing the largest share); 7 of 18 (39\%) are corroborated by at least one other model, providing weak but non-trivial cross-model evidence that the discoveries are not artifacts of a single model's idiosyncrasies.

\begin{figure}[t]
\centering
\includegraphics[width=\columnwidth]{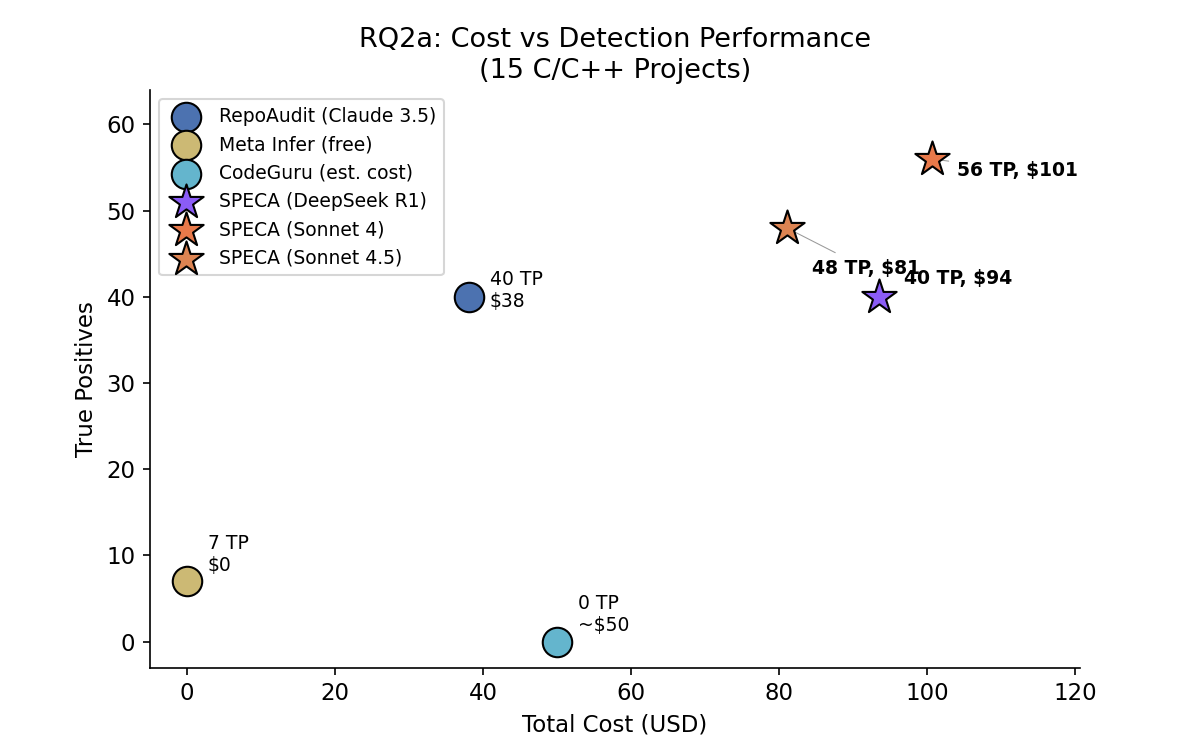}
\Description{Cost vs performance.}
\caption{Cost vs.\ detection performance on the RepoAudit benchmark. At Sonnet~4.5, \sys matches top precision while uniquely contributing author-validated beyond-GT candidates; per-bug cost remains competitive with the best published baseline.}
\label{fig:rq2-cost}
\end{figure}

\noindent\textbf{Property adherence as causal mechanism.}
The per-model count pattern (Sonnet~4: 18; Sonnet~4.5: 12; DeepSeek~R1: 7) is not a simple precision--discovery tradeoff but reflects increasing \emph{property adherence}: more capable models audit more faithfully within the stated property scope, while weaker models drift to related-but-unspecified patterns---some of which turn out to be genuine bugs and others FPs. The empirical signature of this drift is concrete: 5 of 6 externally submitted PRs originate from Sonnet~4's additional (out-of-property-scope) discoveries, not from Sonnet~4.5's tighter set. The engineering implication is that as model capability advances, property-generation quality (Phases~1--3) becomes the binding constraint on coverage---the model will faithfully audit precisely what the properties tell it to audit, and no more. Figure~\ref{fig:rq2-cost} situates \sys's cost--precision position: at Sonnet~4.5, \sys matches the top RepoAudit precision while uniquely contributing double-digit beyond-GT candidates at $\sim\$1.69$/bug.

\noindent\textbf{External validation status.}
We classify validation evidence into four levels of decreasing strength: \textbf{(A)} developer fix commit; \textbf{(B)} maintainer acknowledgement; \textbf{(C)} independent reproduction with documented steps; \textbf{(D)} author analysis only. To upgrade levels, we submitted 8 pull requests across model configurations. Two candidates are externally validated above Level~D: a coturn NULL-deref (PROP-N3-npd-001) at Level~A (fix in later release; PR~\#1841 self-withdrawn after discovering the fix), and an ICU use-after-free race (PROP-U5-uaf-002) at Level~B (maintainer-approved Jira PR~\#3921). Two further PRs were rejected as FPs (ImageMagick PR~\#8628; memcached PR~\#1282), 4 remain open; candidates with incomplete disclosure coordination by camera-ready will be anonymized. Appendix~\ref{app:eval:rq2} reports full per-model cross-corroboration and auxiliary figures.

\section{Discussion}
\label{sec:discussion}

The central finding is that specification-grounded representation changes not just \emph{what} a tool detects but \emph{what its failures mean}: \sys's false positives cluster into three interpretable, phase-traceable categories.

\noindent\textbf{From research artifact to audit pipeline.}
Several engineering choices emerged from live deployment, not offline benchmarking: the proof-attempt framing replaced an adversarial prototype after the latter produced an 88\% FP rate under contest pressure (\S\ref{sec:design:phase5}); the review filter was reduced from five gates to three (\S\ref{sec:design:phase6}) after the dropped gates filtered informational TPs at zero precision; the property neighborhood effect (\S\ref{sec:eval:rq1}), initially unexpected, proved valuable for triage when no single alert was self-validating. The controlled re-experiment in \S\ref{sec:evaluation} is therefore both a methodological isolation step and a record of design choices whose first validation was operational---grounding the claims in deployment, not just offline tuning.

\noindent\textbf{Model capability, property adherence, and asymmetric security.}
\sys benefits disproportionately from model capability: precision climbs from 72.7\% (DeepSeek~R1) to 88.9\% (Sonnet~4.5), a 16.2~pp gain vs.\ 10.5~pp across RepoAudit's configurations. The gain comes with a structural shift: more capable models adhere more closely to the stated property scope, reducing FPs but also reducing beyond-property discoveries (Sonnet~4.5: 12 new, 6~FP; Sonnet~4: 18 new, 13~FP). The implication is that \emph{as models improve, property generation quality becomes the dominant factor in coverage}: the model faithfully verifies what the properties encode but will not compensate for properties the pipeline fails to derive. This makes Phases~1--3 the primary research frontier---a concrete, tractable target in contrast to general code-reasoning improvement. More broadly, this exposes a structural asymmetry: open-ended bug finding favors the attacker, whereas \sys turns LLM capability into a defensive asset by grounding search in explicit specification-derived properties.

\noindent\textbf{Limitations.}
\emph{(i)~Causal attribution:} although RQ2 is partially controlled at the model backbone, architecture and retrieval still differ; results are consistent with a representation effect but do not isolate it.
\emph{(ii)~External validation:} the 12 Sonnet~4.5 candidates were initially Level~D; two cross-model beyond-GT discoveries have been externally validated (Level~A, Level~B), both from Sonnet~4/DeepSeek~R1; two others were maintainer-confirmed FPs.
\emph{(iii)~External validity:} RQ1 covers Ethereum clients and RQ2 a C/C++ benchmark; broader validation across protocol stacks and cryptographic libraries remains future work. Appendix~\ref{app:discussion} discusses additional threats to validity.

\section{Conclusion}
\label{sec:conclusion}

We presented \sys, a specification-anchored audit framework that turns natural-language protocol specifications into a shared property vocabulary for multi-implementation distributed systems. The expert-augmented configuration recovers all 15 H/M/L Sherlock vulnerabilities (automated-only: 8/15) and discovers 4 fix-confirmed bugs across 10 audit targets spanning 6 languages; cluster-level precision (48.7\%) nearly doubles the finding-level rate, and all 15 detecting properties encode specification-derived information unavailable from code alone. On RepoAudit, \sys matches the best baseline precision (88.9\%) and surfaces 12 author-validated beyond-GT candidates with Sonnet~4.5, with two cross-model discoveries externally validated (Level~A, Level~B). All false positives trace to three pipeline-specific root causes, identifying property-generation quality---especially for cryptographic invariants and protocol-lifecycle rules---as the binding constraint and the most tractable extension of this work.

\section*{Ethical Considerations}
\label{sec:ethics}

\noindent\textbf{Responsible disclosure.}
All Sherlock findings correspond to already-adjudicated contest issues.
The 4 novel bugs have been remediated upstream (we reference only public fix commits).
The RepoAudit candidate bugs follow responsible disclosure: 8 pull requests have been
submitted to upstream repositories, 2 candidates externally validated above author review
(one Level~A, fix confirmed in a later release; one Level~B, maintainer Jira approval),
and 2 confirmed as false positives; any unresolved candidate will be anonymized in the
camera-ready version.

\noindent\textbf{Dual-use considerations.}
\sys is a defensive tool, but vulnerability-finding is inherently dual-use.
We believe publication benefits outweigh risks: the compute budget is already accessible
to well-resourced attackers, every finding is traceable to an explicit property (aiding
governance), and the pipeline is most useful in specification-grounded contexts where
defenders operate.

\noindent\textbf{Human subjects and compute cost.}
This work involves no human subjects; we used only publicly released contest
adjudication data.
Total API cost was \$400--620 (RQ1) and \$81--101 per model (RQ2), reported transparently
in \S\ref{sec:eval:setup}.

\bibliographystyle{IEEEtran}
\bibliography{references}

\section*{Open Science}
\label{sec:openscience}

In accordance with the NDSS 2027 open science guidance, we enumerate the artifacts
supporting this work, their availability, and justifications for any restrictions.

\noindent\textbf{Artifacts provided for review.}
The following artifacts are available at
\url{https://anonymous.4open.science/r/speca}
and via the submission system:
(i)~the evaluation scripts (\texttt{evaluate.py}, \texttt{matchers.py},
\texttt{analyze\_deep.py});
(ii)~all raw result data (\texttt{evaluation\_summary.json},
\texttt{findings\_labels.csv}, \texttt{deep\_analysis.json}) and generated figures;
(iii)~per-repository output directories for all 10 Sherlock clients and all 15
RepoAudit C/C++ projects;
(iv)~all intermediate phase outputs (specification indices, subgraphs, generated
properties, pre-/post-review findings) enabling offline inspection and partial
re-execution; and
(v)~the \sys pipeline implementation, including orchestration code, worker prompts,
schema definitions, the STRIDE+CWE property generation templates, and the
severity-preserving review filter.

\noindent\textbf{Artifacts not publicly released and justification.}
The pipeline source code and prompt templates are provided to reviewers via the
anonymous artifact but will not be released as open-source software.
This decision reflects dual-use risk: the pipeline constitutes a vulnerability-finding
capability that, if publicly released, could lower the cost of offensive exploitation
against specification-governed systems.
While the per-bug cost is already within reach of well-resourced attackers
(\S\ref{sec:eval:setup}), broad public release would remove the barrier for less
sophisticated adversaries.
Additionally, responsible disclosure for the RepoAudit candidate bugs is in progress
(8 PRs submitted, 2 maintainer-confirmed, 2 confirmed FP, 4 awaiting response); raw
LLM responses referencing unfixed vulnerabilities are redacted from the artifact and
will remain so until disclosure is complete.

\noindent\textbf{Reproducibility.}
\sys's pipeline is deterministic given a fixed model configuration and seed, except
for residual nondeterminism in provider-side LLM responses.
We report exact model versions (Claude Opus, Claude Sonnet~4.5, Sonnet~4, DeepSeek~R1)
and total token consumption per phase to enable cost estimation prior to replication.
The artifact includes all intermediate phase outputs so that reviewers can inspect
the full pipeline trace and re-run individual phases offline without API access.

\noindent\textbf{Data provenance.}
The Sherlock contest data is sourced from the publicly released Sherlock Ethereum
Fusaka contest adjudication record; we use only publicly released contest issues and
severity judgments.
The RepoAudit benchmark data is sourced from the published RepoAudit
artifact~\cite{repoaudit}; we use only the 15 open-source C/C++ projects included in
that benchmark.

\appendix

\section{Extended Related Work}
\label{app:related}

This appendix supplements Section~\ref{sec:related} with additional differentiation
along the same three-axis organization.

\noindent\textbf{LLM-augmented code-level vulnerability detection.}
Recent work pairs LLMs with classical static analysis or graph
representations to improve taint-style and path-sensitive bug
detection: LLift~\cite{llift}, BugLens~\cite{buglens}, IRIS~\cite{iris},
LLMxCPG~\cite{llmxcpg}, and the autonomous repository-scale agent
RepoAudit~\cite{repoaudit}; GoSonar~\cite{gosonar} extends this to logical
bugs in memory-safe languages. Empirical studies further show that LLMs
exhibit non-determinism and reasoning errors on this
task~\cite{ullahsp24} and that popular benchmarks overstate model
performance~\cite{primevul}. All of these methods start from
repository-local information; \sys instead derives security properties
from specifications before reading code, addressing correctness
conditions that are not visible at the code level alone.

\noindent\textbf{LLM-driven security property generation.}
PropertyGPT~\cite{propertygpt} (NDSS 2025 Distinguished Paper) applies
retrieval-augmented generation to transfer human-written properties to new
smart contracts; GPTScan~\cite{gptscan} and iAudit~\cite{iaudit} target
smart-contract logic vulnerabilities by decomposing them into scenarios
and properties; FormalBench~\cite{formalbench} evaluates LLM ability to
synthesize formal program specifications and finds that multi-phase
pipelines with structured review are necessary for reliable inference.
These works share \sys's emphasis on explicit properties as the unit of
analysis but derive properties from code or pre-existing verified
properties rather than from authoritative natural-language
specifications, and target predominantly smart contracts rather than
multi-implementation distributed systems.

\noindent\textbf{Specification-driven security analysis.}
RFCAudit~\cite{rfcaudit} audits protocol implementations against RFCs
using an autonomous LLM agent (47 functional bugs at 81.9\% precision);
EDEFuzz~\cite{edefuzz} uses API specs to test for data-exposure
vulnerabilities; Jin et al.~\cite{jinase25} show that complex prompting
paradoxically raises false-negative rates due to overcorrection bias---a
finding that motivates \sys's severity-preserving filter rather than
aggressive LLM post-processing; Cortier et al.~\cite{cortiereprintideal}
show that even formal specifications can miss intuitively expected
security properties, paralleling \sys's empirical observation that many
false positives stem from specification-interpretation mismatches.
QUACK~\cite{quack} and ISLAB~\cite{islab} demonstrate property-based
security enforcement that \sys extends from enforcement to auditing.
Wadhwa et al.~\cite{nayakccs24} formalize specification--implementation
gaps in blockchain ordering protocols, confirming that such gaps are an
actively studied problem in the multi-implementation setting \sys
targets. Among these, RFCAudit is closest in spirit; \sys differs in
three respects beyond the functional-correctness vs.\ security-property
axis: it derives explicit, typed properties (vs.\ implicit RFC
compliance), supports controlled cross-implementation comparison under a
shared property vocabulary, and traces every false positive to a
specific pipeline phase.

\noindent\textbf{Evaluation methodology considerations.}
Arp et al.~\cite{arppierazzi} catalog common pitfalls in ML-based
security evaluation (sampling bias, label leakage, unrealistic
distributions) that inflate reported performance. Our design mitigates
these: the Sherlock benchmark uses independently adjudicated real-world
contest data, precision is reported at three granularities, and the
structured false-positive taxonomy adds failure-mode analysis beyond
aggregate metrics.

\section{Extended Design Details}
\label{app:design}

\subsection{Phase 4: code pre-resolution details}

Phase~4 uses Tree-sitter-based symbol analysis (via \texttt{get\_symbols} and
\texttt{run\_query}) to link each property to concrete source locations.
It identifies the primary enforcement location by matching specification concepts
(function names, data structures, protocol messages) to source code symbols,
discovers callers and callees to establish the relevant code neighborhood, and builds
a subgraph index (\texttt{01b\_SUBGRAPH\_INDEX.json}) mapping specification-level
concepts to implementation artifacts.
Properties classified as Informational severity are dropped at this stage.
Pre-resolution decouples code discovery from security reasoning; an alternative would
be to let Phase~5 search for relevant code on-the-fly, but on-the-fly search is
expensive and produces inconsistent code neighborhoods across properties.
Pre-resolution reduces downstream audit token consumption by 40--60\%.
The trade-off is that pre-resolution may miss code locations unreachable through symbol
analysis.

\subsection{Cross-implementation end-to-end example}

Consider auditing a property requiring nodes to rotate custody columns every
\texttt{CUSTODY\_ROTATION\_INTERVAL} epochs across two implementations.
Phase~3 generates the property from the specification with a unique identifier and
invariant classification.
Phase~4 resolves this property to \texttt{rotateCustody()} in Implementation~A's Nim
source and to \texttt{handleCustodyRotation()} in Implementation~B's Go source.
Phase~5 audits both: in Implementation~A, the interval check uses \texttt{>=} instead of
\texttt{==}, causing premature rotation (finding: Low severity); in Implementation~B,
the function is called correctly (finding: PASS).
Phase~6 confirms Implementation~A's finding as a confirmed vulnerability (the rotation
interval is controlled by protocol parameters, not trusted input).
The same property produces a true positive in one implementation and a pass in
another---exactly the kind of divergence that a shared property vocabulary is designed
to reveal.

\section{Extended Evaluation: RQ1}
\label{app:eval}

\subsection{Experimental setup details}
\label{app:eval:setup}

\begin{table}[h]
\centering
\caption{Finding label definitions. Precision $=$ non-FP rate: all labels except \texttt{fp\_invalid} and \texttt{fp\_review} count as non-FP. *\texttt{potential-info} findings are plausible security observations below the threshold for confirmed vulnerability.}
\label{tab:label-defs}
\footnotesize
\setlength{\tabcolsep}{4pt}
\begin{tabular}{l>{\raggedright\arraybackslash}p{4.2cm}c}
\toprule
\textbf{Label} & \textbf{Definition} & \textbf{TP?} \\
\midrule
\texttt{tp} & Matches valid H/M/L contest issue & Yes \\
\texttt{tp\_info} & Informational-severity or real low-severity & Yes \\
\texttt{fixed} & Independently fixed on target repository & Yes \\
\texttt{partially\_fixed} & Partially addressed & Yes \\
\texttt{potential-info} & Plausible but unconfirmed & Yes* \\
\texttt{fp\_invalid} & False positive (invalid reasoning) & No \\
\texttt{fp\_review} & False positive (review-gate error) & No \\
\bottomrule
\end{tabular}
\end{table}

Table~\ref{tab:label-defs} provides the complete finding-label definitions referenced from the main body. Below we give a worked matching example.

\noindent\textbf{Matching example.}
Contest issue~\#308 reports a missing validation of data column index bounds in gossip
message handling.
\sys produces four findings for the same repositories: finding
\texttt{PROP-6a4369e9-inv-009} (Lighthouse), ``Column index must be validated against
\texttt{NUMBER\_OF\_COLUMNS} before acceptance''---\textbf{match accepted}, same root
cause and same impact; \texttt{PROP-6a4369e9-inv-010} (Lighthouse), ``EL-sourced custody
columns skip KZG verification''---\textbf{partial match, excluded} (different root
cause); and \texttt{PROP-6a4369e9-inv-008} (Nimbus), ``Unreachable code path in data
column sidecar construction''---\textbf{no match}.
Under our protocol, issue~\#308 is counted as recovered (at least one match), and
\texttt{PROP-6a4369e9-inv-009} counts as one true positive.

\subsection{Auxiliary RQ1 figures and tables}
\label{app:eval:accounting}

\begin{figure}[h]
\centering
\includegraphics[width=\columnwidth]{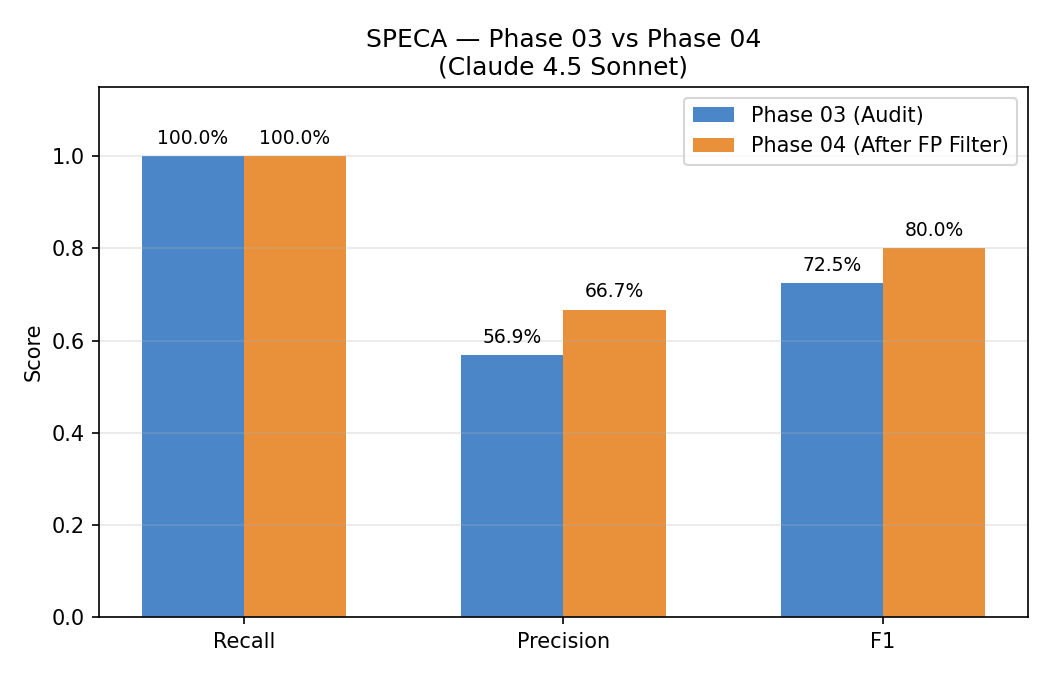}
\Description{Phase 5 vs Phase 6 comparison.}
\caption{Phase~5 (Audit) vs Phase~6 (FP Filter). Phase~6 improves precision from 56.9\% to 66.7\% while preserving 100\% recall on H/M/L findings, raising F1 from 72.5\% to 80.0\%.}
\label{fig:rq1-phase}
\end{figure}

\begin{figure}[h]
\centering
\includegraphics[width=\columnwidth]{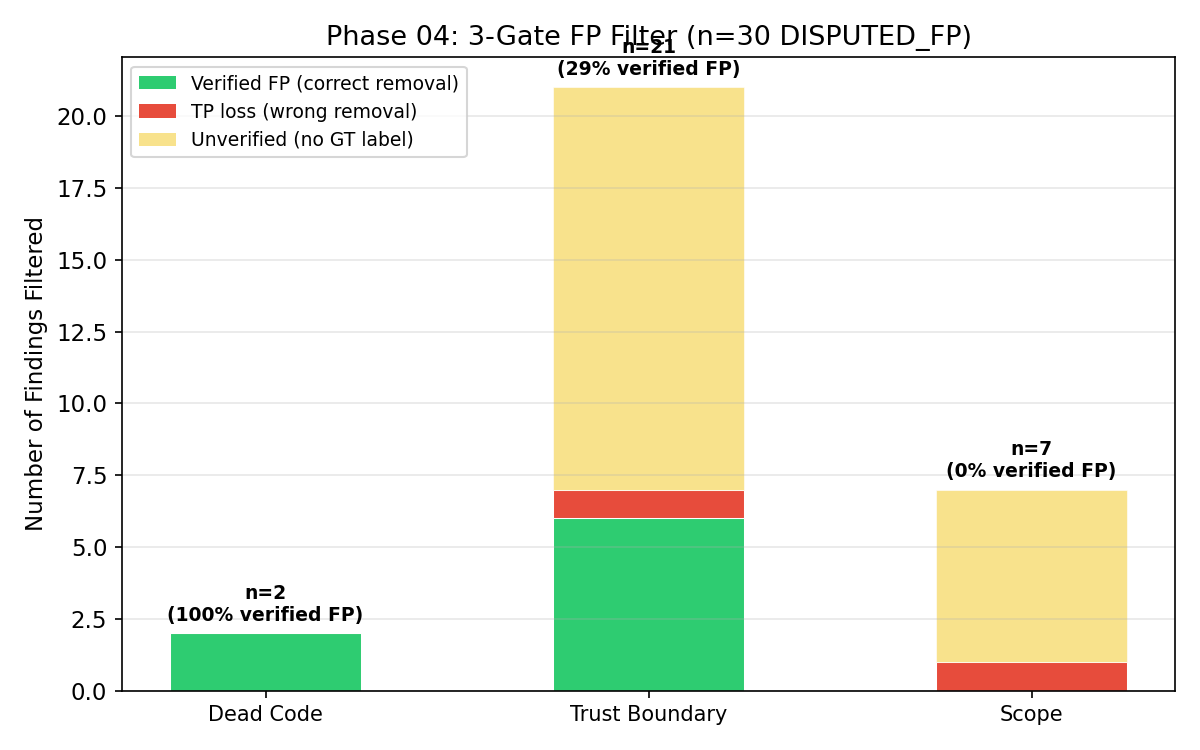}
\Description{Gate effectiveness.}
\caption{Phase 6 three-gate FP filter effectiveness (N=30 DISPUTED\_FP).}
\label{fig:rq1-gates}
\end{figure}

\begin{figure}[h]
\centering
\includegraphics[width=\columnwidth]{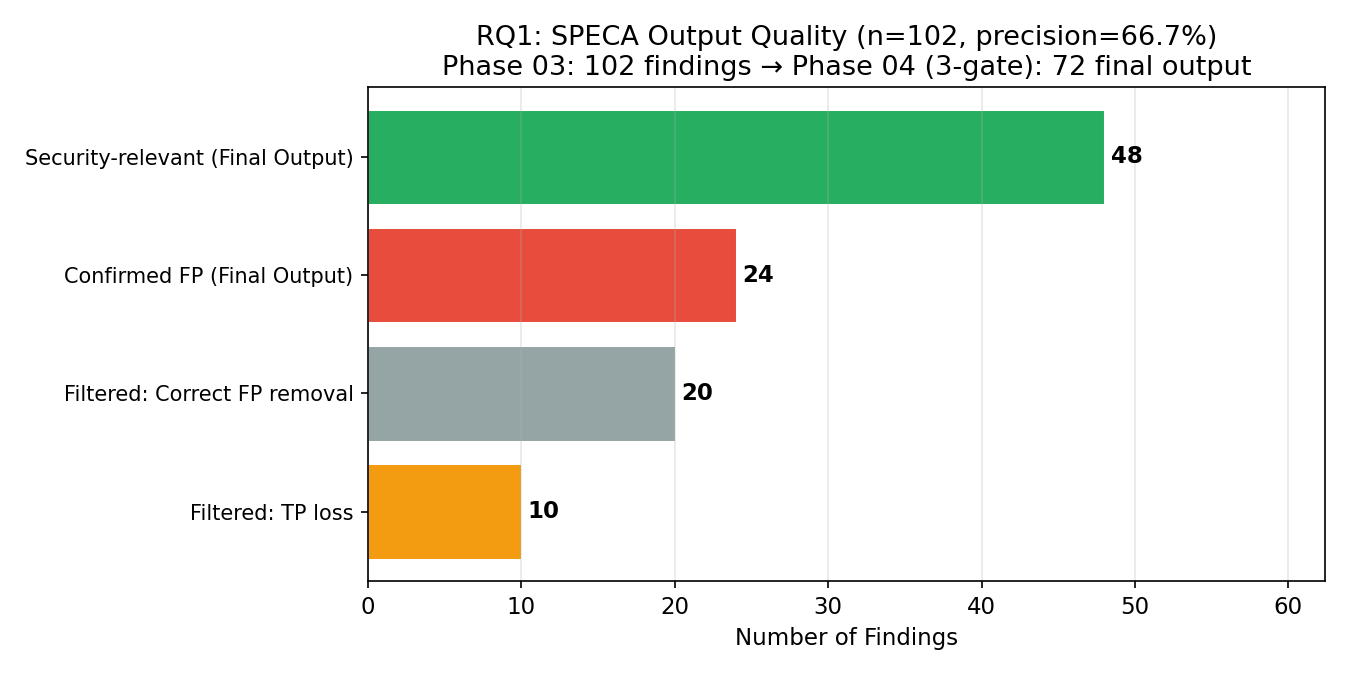}
\Description{Output quality breakdown bar chart.}
\caption{\sys output quality breakdown (N=102). Of 102 pre-review findings, 48 are security-relevant in the final output, 24 are confirmed FPs, 20 are correctly filtered by Phase 6, and 10 are TPs lost to Phase 6 filtering.}
\label{fig:rq1-labels}
\end{figure}

Table~\ref{tab:rq1-accounting} provides a full accounting reconciliation of all finding counts reported in RQ1; Figure~\ref{fig:rq1-labels} provides the output-quality breakdown.

\begin{table}[h]
\centering
\caption{Accounting reconciliation of all finding counts reported in RQ1. Each count traces to a single partition so that the precision numerator/denominator and FP-analysis population can be verified by a reader.}
\label{tab:rq1-accounting}
\footnotesize
\setlength{\tabcolsep}{4pt}
\begin{tabular}{lrl}
\toprule
\textbf{Stage} & \textbf{Findings} & \textbf{Source} \\
\midrule
Phase 5 total output & 647 & All properties audited \\
\quad Not-a-vuln / out-of-scope & 545 & Proof succeeded \\
\quad Positive output & \textbf{102} & Flagged findings \\
\midrule
\multicolumn{3}{l}{\emph{Pre-review (N=102):}} \\
\quad Non-FP & \textbf{58} & 19+5+26+2+6 \\
\quad FP & \textbf{44} & 40+4 \\
\midrule
\multicolumn{3}{l}{\emph{Phase 6 review of 102:}} \\
\quad Survived & \textbf{72} & \\
\quad DISPUTED\_FP & \textbf{30} & 20 FP, 10 TP \\
\midrule
\multicolumn{3}{l}{\emph{Post-review (N=72):}} \\
\quad Non-FP & \textbf{48} & 19+18+5+1+5 \\
\quad FP & \textbf{24} & 23+1 \\
\quad Confirmed-useful & \textbf{43} & 19+18+5+1 \\
\midrule
\multicolumn{3}{l}{\emph{FP analysis universes:}} \\
\quad Total FPs (both phases) & \textbf{44} & 40+4 \\
\quad Deeply analyzed FPs & \textbf{16} & 8 post + 8 filtered \\
Property-type ablation subset & \textbf{88} & 31+7+50 \\
\bottomrule
\end{tabular}
\end{table}

\subsection{Property neighborhoods and cluster-level evaluation}
\label{app:eval:neighborhoods}

Many issues are recovered not by a single alert but by multiple complementary
properties that form a \emph{property neighborhood} around the root cause.
The deep analysis identifies 11 issues matched by 2 or more properties (e.g., issues
\#308 and \#331 each matched by 4 properties, issue \#169 by 3).
Conversely, 2 properties each match multiple issues (e.g.,
\texttt{PROP-6a4369e9-inv-047} matches both \#109 and \#15).

The main body Table~\ref{tab:rq1-cluster} reports the finding-level vs
cluster-level comparison.
At the cluster level, 39 distinct root-cause clusters produce 19 clusters matching H/M/L
issues (some issues are detected across multiple repositories, producing separate
clusters per repository).
The cluster-level strict precision (48.7\%) is substantially higher than finding-level
strict precision (26.4\%), confirming that the property neighborhood provides genuine
redundancy rather than alert duplication.

\subsection{Cross-implementation patterns}
\label{app:eval:crossimpl}

Eleven issues appear across multiple repositories under the same property vocabulary
(e.g., issue~\#308 appears in Grandine, Lighthouse, Nimbus, and Prysm; \#169 in
Grandine, Lodestar, and Prysm; \#331 in Lighthouse, Lodestar, and Prysm).
Property \texttt{PROP-6a4369e9-inv-014} asserts that blob parameters at fork boundaries
must be computed from the specification's canonical schedule.
Under the shared property vocabulary, this property produces divergent outcomes:
Grandine, Lighthouse, and Prysm use a local config default that diverges from the
specification at fork epochs (all three TP, matched to issue~\#11), while Nimbus
correctly reads the canonical schedule and the property passes---illustrating how a
single property can function simultaneously as a true-positive detector in deviating
implementations and a conformance certificate in compliant ones.

\begin{figure}[h]
\centering
\includegraphics[width=\columnwidth]{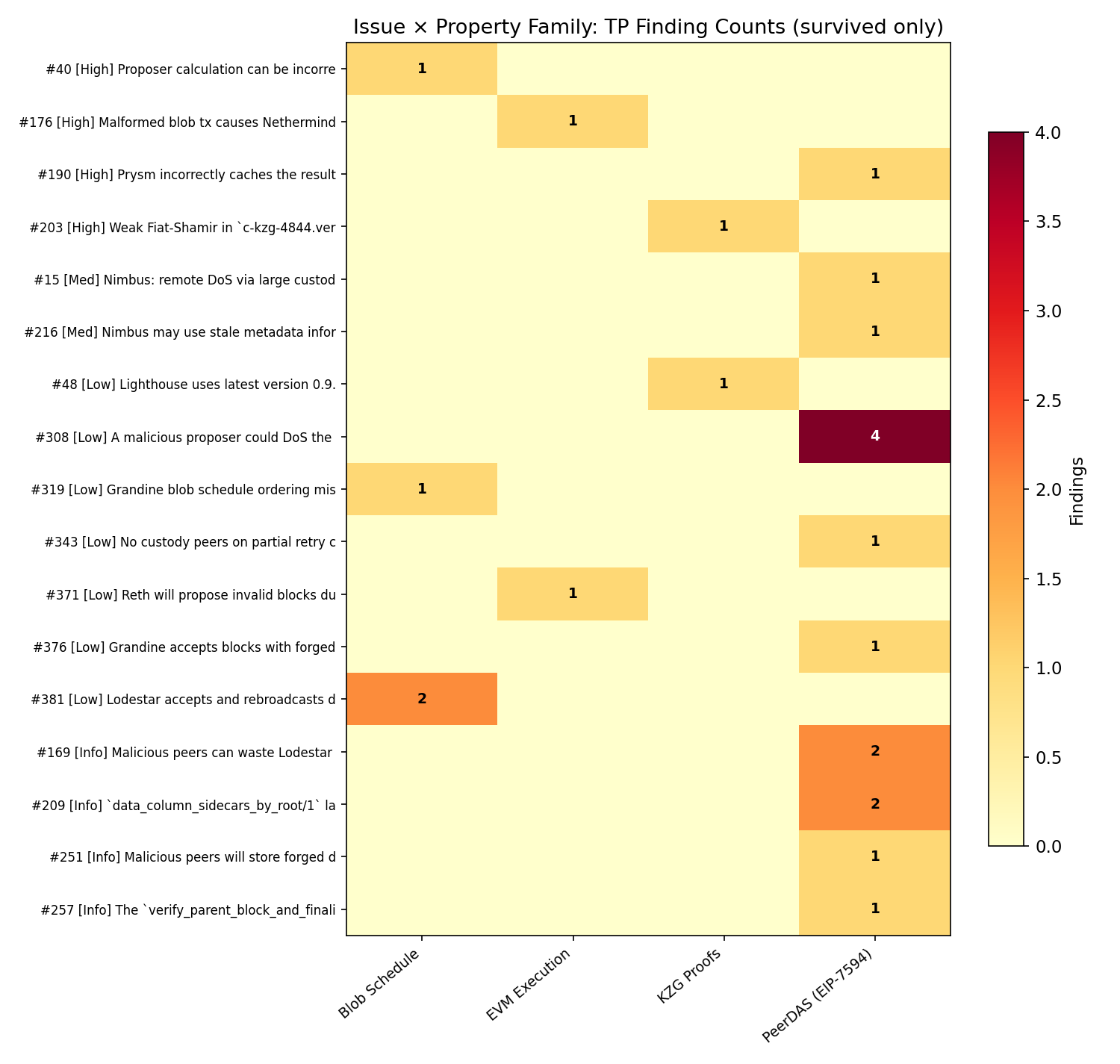}
\Description{Heatmap of TP findings by issue and property family.}
\caption{Issue $\times$ property family TP finding counts (post-review). The property neighborhood effect is visible as horizontal bands where a single issue is detected by multiple complementary properties; shared vertical structure across repositories reflects the cross-implementation comparability the property vocabulary enables.}
\label{fig:rq1-heatmap}
\end{figure}

Figure~\ref{fig:rq1-perrepo} (main body) shows the per-repository finding distribution.

\subsection{False-positive analysis: full detail}
\label{app:eval:fp}

\begin{figure}[h]
\centering
\includegraphics[width=\columnwidth]{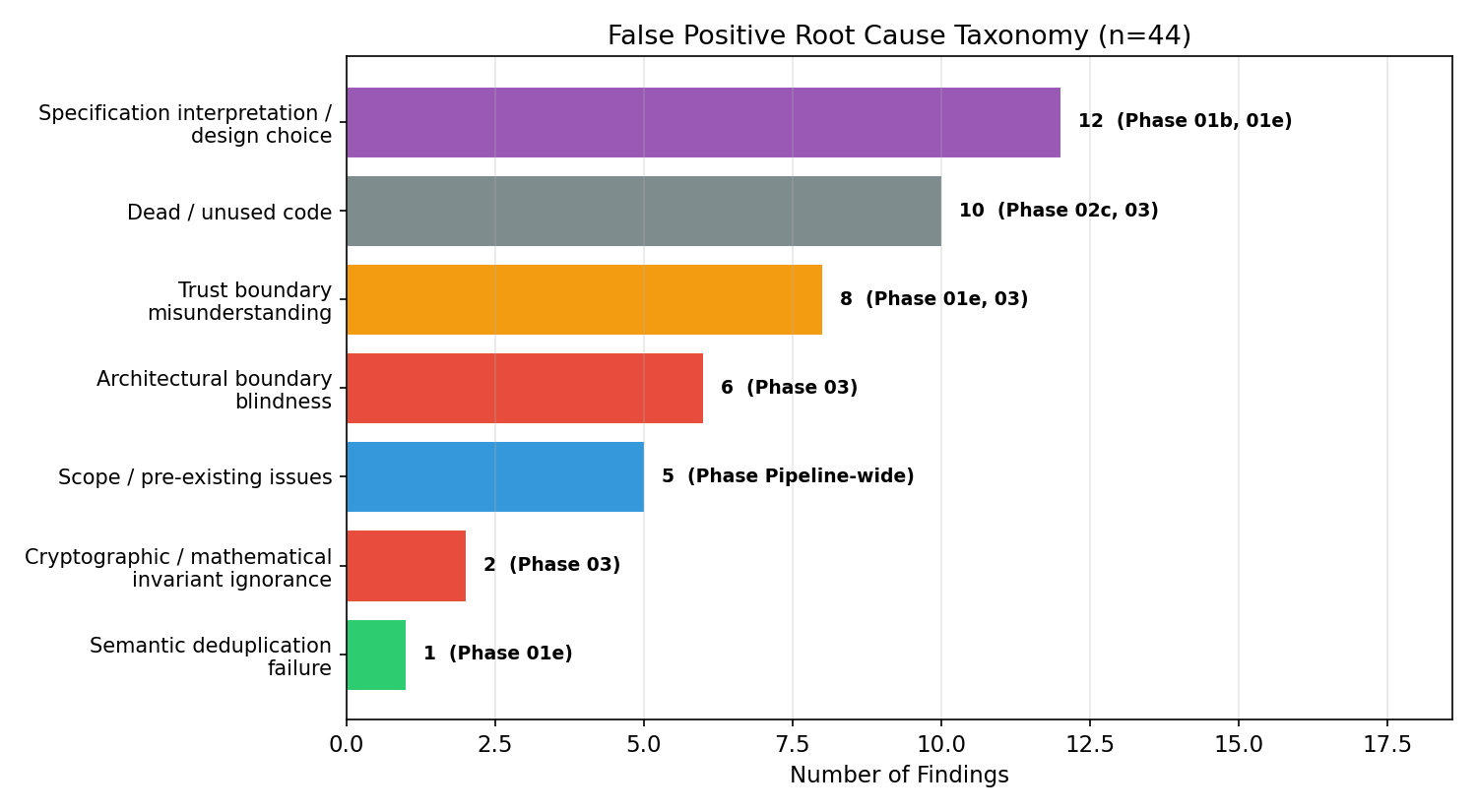}
\Description{FP taxonomy by phase origin.}
\caption{False-positive root-cause taxonomy with pipeline-phase origins (N=16 deeply analyzed from 44 total FPs; see Table~\ref{tab:fp-taxonomy}). The three causes each map to a distinct pipeline phase, making the failure surface structurally analyzable.}
\label{fig:rq1-fp-tax}
\end{figure}

Figure~\ref{fig:rq1-fp-tax} visualizes the false-positive root-cause taxonomy.

Of the 16 deeply analyzed FPs, trust-boundary errors account for 8 findings, code
reading errors for 6, and specification misinterpretation for 2.
Phase~6 catches half of trust-boundary FPs (4/8) and two-thirds of code-reading FPs
(4/6), but none of the specification-misinterpretation FPs (0/2), because the
mechanical gates cannot resolve semantic ambiguity.
Each root cause maps to a specific, implementable improvement:
adding explicit trust boundary configuration to Phase~3 would address trust-boundary
FPs; improving Phase~5's code reading through richer context windows or multi-pass
verification would address code reading errors; and requiring Phase~5 to re-read the
relevant specification section before classifying a deviation as a vulnerability would
address specification-misinterpretation FPs.

\subsection{Property-type ablation: supporting detail}
\label{app:eval:ablation}

\begin{figure}[h]
\centering
\includegraphics[width=\columnwidth]{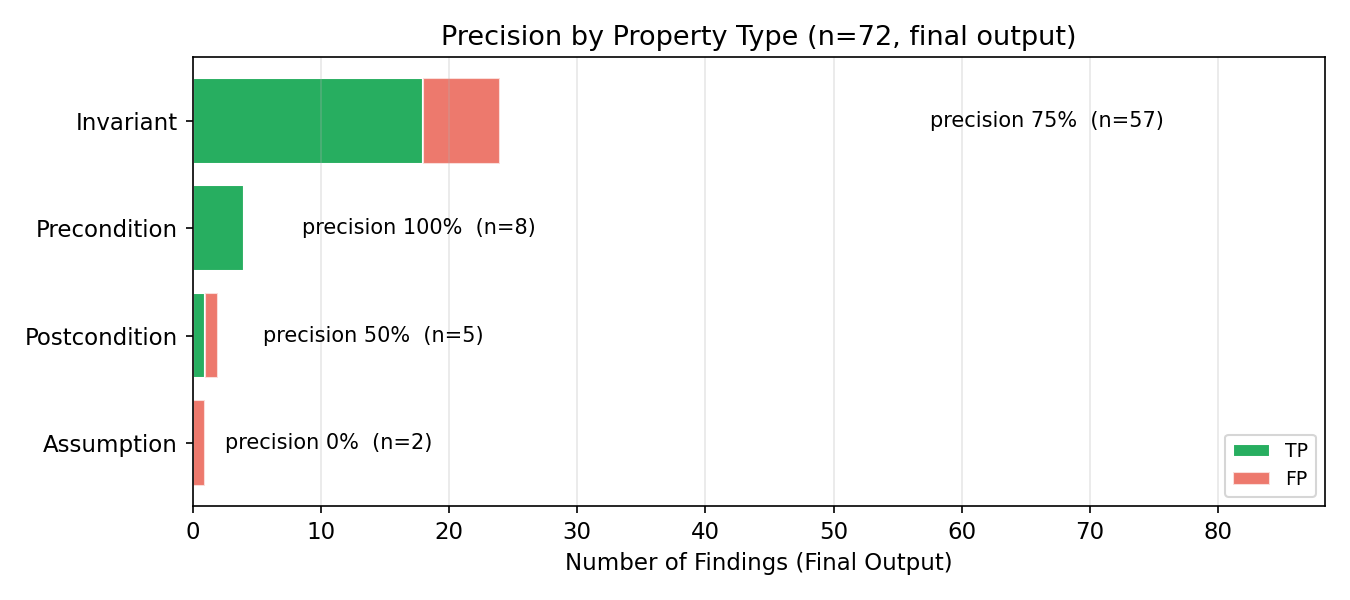}
\Description{Precision by property type.}
\caption{Precision by property type (final output).}
\label{fig:rq1-ablation}
\end{figure}

The main body Table~\ref{tab:prop-ablation} reports precision by property type over
the 88-finding ablation subset.
Figure~\ref{fig:rq1-ablation} visualizes the per-type precision; the effect of
excluding assumption-type findings on the post-review universe is shown in
Table~\ref{tab:rq1-no-asm}.

\begin{table}[h]
\centering
\caption{Effect of excluding assumption properties on the post-review universe (N=72).}
\label{tab:rq1-no-asm}
\small
\begin{tabular}{lrr}
\toprule
\textbf{Metric} & \textbf{All types} & \textbf{Excl.\ asm} \\
\midrule
Precision & 66.7\% & 67.1\% \\
H/M/L match rate & 26.4\% & 27.1\% \\
\bottomrule
\end{tabular}
\end{table}

\subsection{Specification contribution analysis: supporting detail}
\label{app:eval:spec}

The main body Table~\ref{tab:spec-essential} and \S\ref{sec:eval:rq1-spec} summarize
the retrospective spec-essentiality classification. For each detecting property, the
provenance chain is: property $\rightarrow$ subgraph (\texttt{SG-NNN}) $\rightarrow$
specification section $\rightarrow$ \texttt{INV} label.
For issues \#319 and \#376, the complete chain terminates at a specific \texttt{INV}
label; for the remaining 13, the chain establishes spec provenance through subgraph
references.

\section{Extended Evaluation: RQ2}
\label{app:eval:rq2}

\begin{table*}[h]
\centering
\caption{Eighteen author-validated candidate bugs identified by \sys (Sonnet~4) beyond the RepoAudit ground truth. Two candidates have been externally validated above Level~D: one at Level~A (fix confirmed in later release) and one at Level~B (maintainer Jira approval). Cross-model corroboration: 7 of 18 (39\%) detected by at least one other model configuration.}
\label{tab:new-bugs}
\footnotesize
\setlength{\tabcolsep}{4pt}
\begin{tabular}{r l l c >{\raggedright\arraybackslash}p{2.55in} c >{\raggedright\arraybackslash}p{1.25in}}
\toprule
\textbf{\#} & \textbf{Property ID} & \textbf{Project} & \textbf{Type} & \textbf{Description} & \textbf{Val.} & \textbf{Evidence} \\
\midrule
1 & PROP-M1-mlk-014 & libsass & MLK & Memory leak in CSS AST processing on error path & D & Author review \\
2 & PROP-M4-mlk-001 & linux/sound & MLK & Resource leak in audio graph setup on error path & D & Author review \\
3 & PROP-M4-mlk-009 & linux/sound & MLK & q6apm\_dai\_open: graph resource leak on constraints & D & Author review \\
4 & PROP-M5-mlk-003 & linux/mm & MLK & DAMON resource leak on reclaim error path & D & Author review \\
5 & PROP-M5-mlk-004 & linux/mm & MLK & DAMON target allocation leak on setup failure & D & Author review \\
6 & PROP-M5-mlk-007 & linux/mm & MLK & DAMON scheme resource leak on init error & D & Author review \\
7 & PROP-N1-npd-003 & sofa-pbrpc & NPD & ResolveAddress: NULL-deref on DNS resolution failure & D & PR \#251 (awaiting) \\
8 & PROP-N3-npd-001 & coturn & NPD & allocation\_get\_new\_ch\_info: NULL channel-map deref & \textbf{A} & Fix in later release \\
9 & PROP-N4-npd-005 & libfreenect & NPD & registration\_table: NULL-deref on malloc failure & D & PR \#697 (awaiting) \\
10 & PROP-U1-uaf-004 & Redis & UAF & Use-after-free in benchmark client error path & D & Author review \\
11 & PROP-U2-uaf-004 & linux/drivers/peci & UAF & Use-after-free in PECI driver error path & D & Author review \\
12 & PROP-U2-uaf-007 & linux/drivers/peci & UAF & Use-after-free in PECI auxiliary device cleanup & D & Author review \\
13 & PROP-U3-uaf-002 & shadowsocks-libev & UAF & Use-after-free in server connection teardown & D & Author review \\
14 & PROP-U3-uaf-003 & shadowsocks-libev & UAF & Use-after-free in UDP relay cleanup & D & Author review \\
15 & PROP-U3-uaf-007 & shadowsocks-libev & UAF & Use-after-free in tunnel connection handler & D & Author review \\
16 & PROP-U3-uaf-008 & shadowsocks-libev & UAF & Use-after-free in connection cleanup path & D & Author review \\
17 & PROP-U5-uaf-002 & icu/i18n & UAF & collationroot cleanup: thread-safety race condition & \textbf{B} & Jira approved (PR \#3921) \\
18 & PROP-U5-uaf-003 & icu/i18n & UAF & Region cleanupRegionData: double-delete on shutdown & D & PR \#3922 (Jira pending) \\
\bottomrule
\end{tabular}
\end{table*}

\begin{figure}[h]
\centering
\includegraphics[width=\columnwidth]{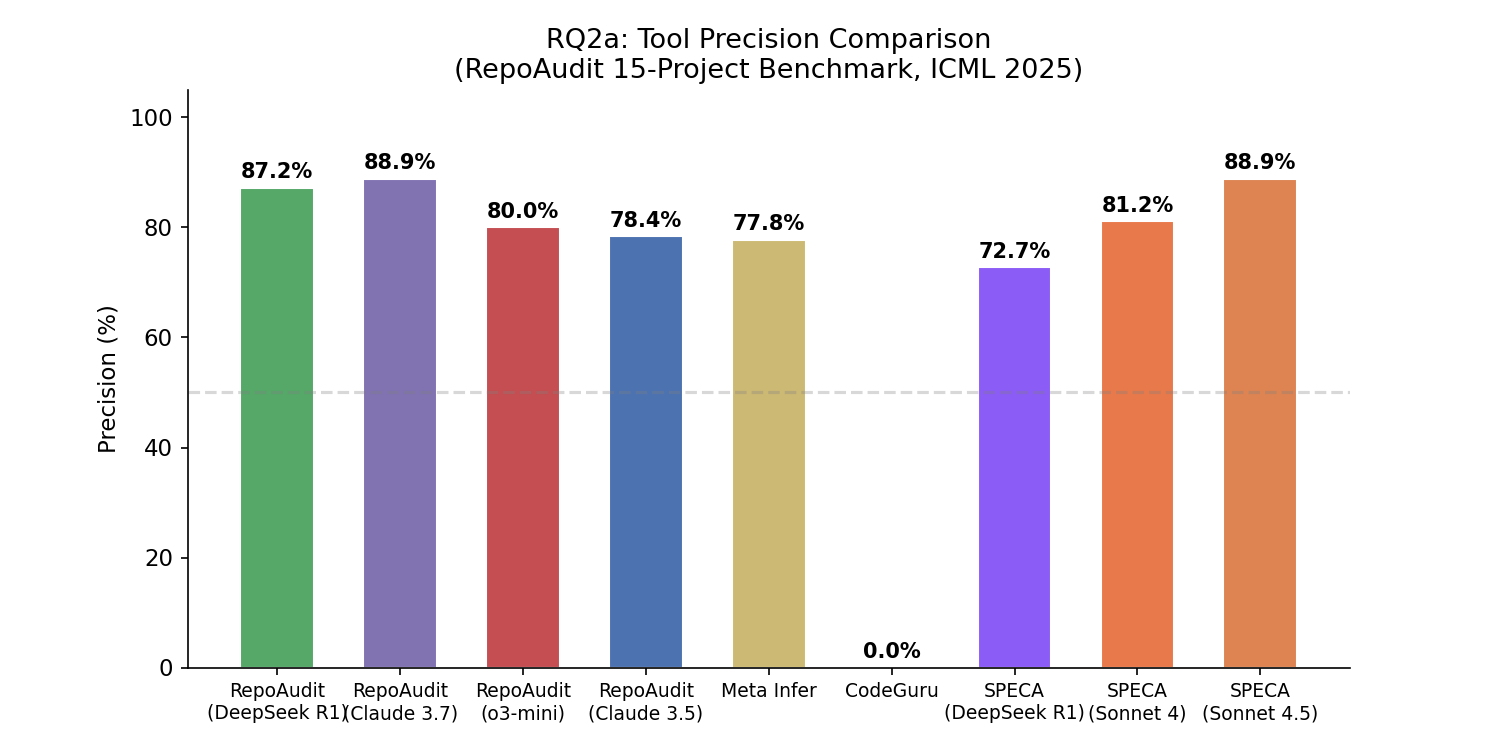}
\Description{Precision comparison across tools.}
\caption{Tool precision on the RepoAudit benchmark. \sys (Sonnet 4.5) matches the best RepoAudit configuration (88.9\%) while surfacing 12 author-validated candidate bugs beyond the established ground truth.}
\label{fig:rq2-prec}
\end{figure}

\begin{figure}[h]
\centering
\includegraphics[width=\columnwidth]{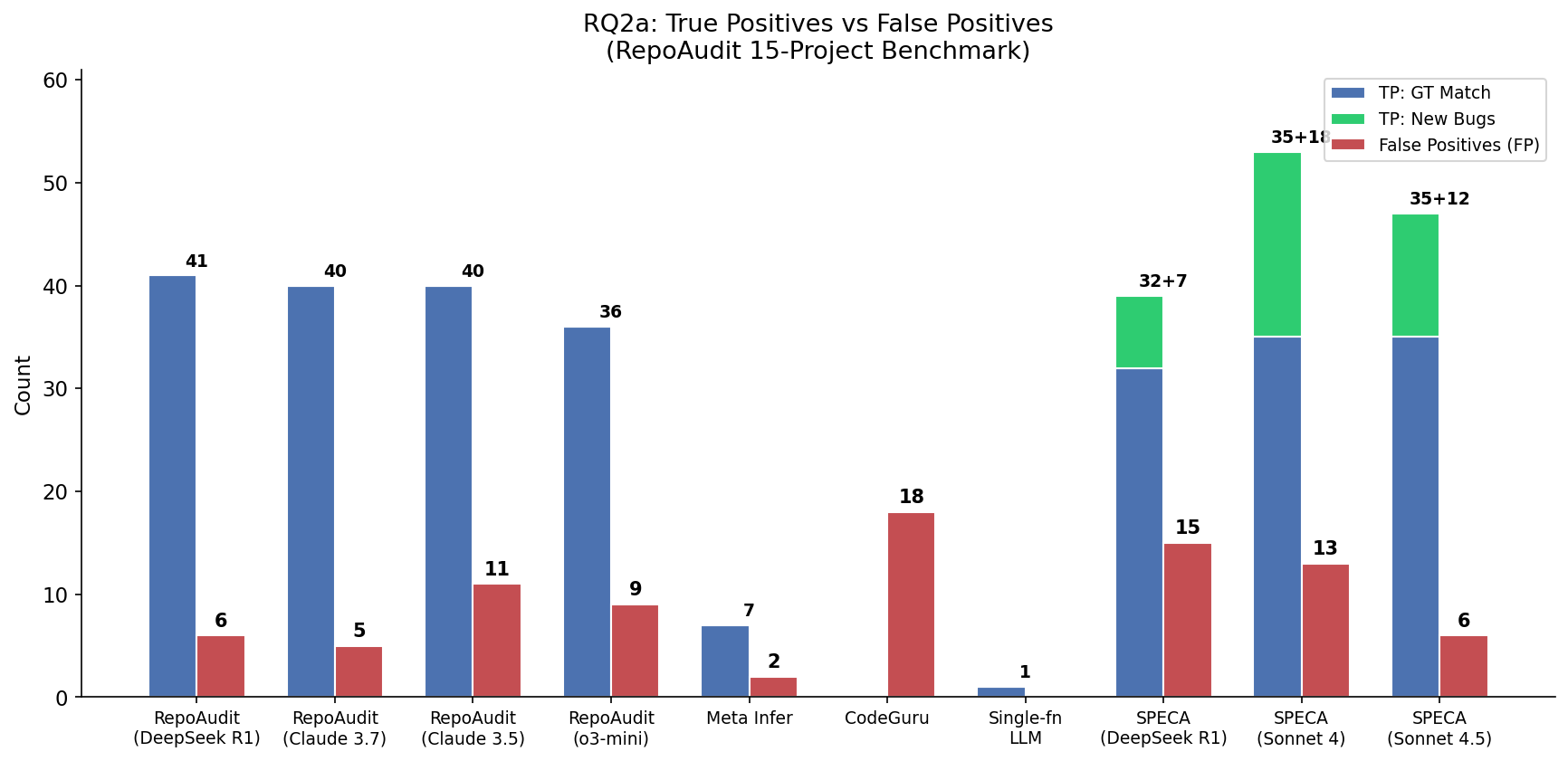}
\Description{True vs false positives.}
\caption{True positives vs false positives across all tools.}
\label{fig:rq2-tpfp}
\end{figure}

\begin{figure}[h]
\centering
\includegraphics[width=\columnwidth]{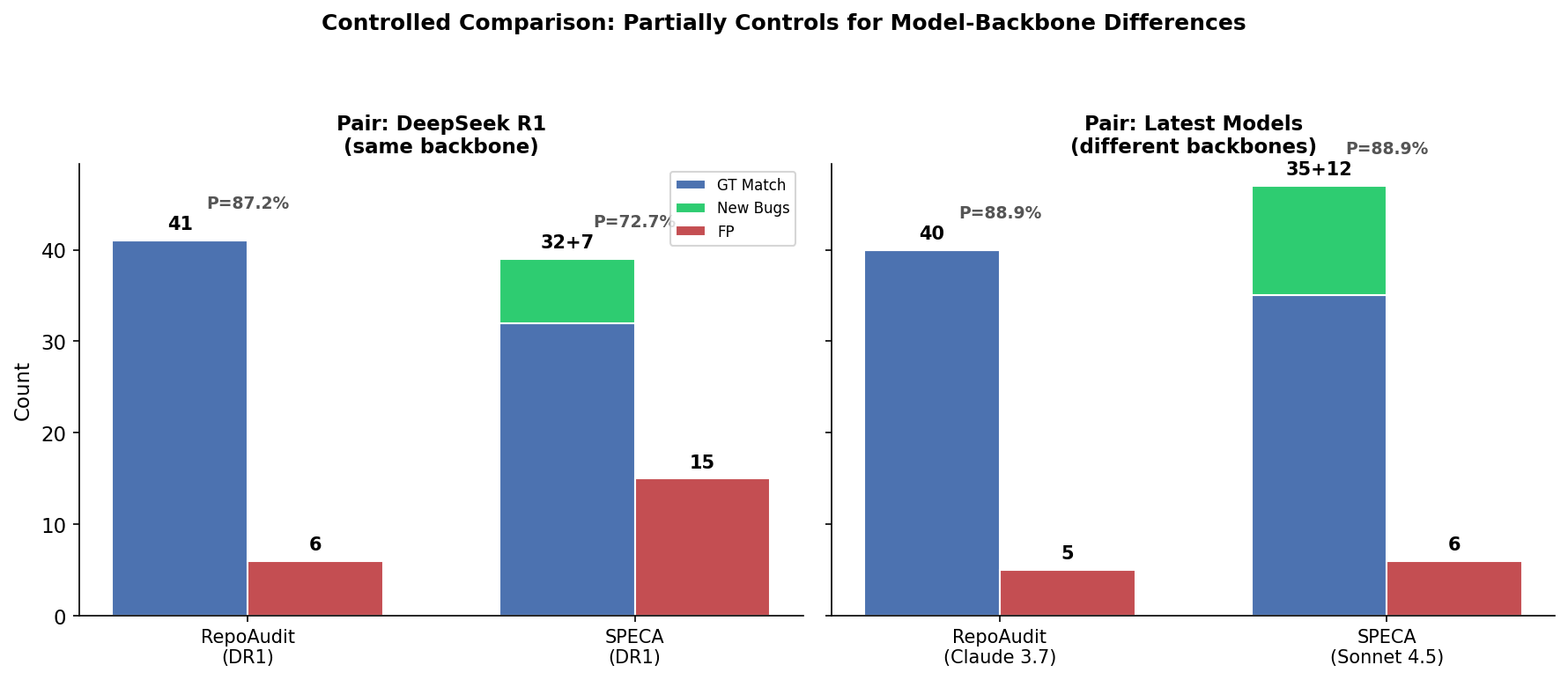}
\Description{Symmetric comparison.}
\caption{Partially controlled comparison across model backbones. Left: same-backbone (DeepSeek~R1) isolates the representation effect under matched model capability. Right: latest-models comparison shows \sys matching the best baseline precision while uniquely contributing beyond-GT candidates.}
\label{fig:rq2-sym}
\end{figure}

\subsection{Cost--output trade-off}

\sys's cost (\$400--620 for RQ1, \$81.05 for RQ2) is approximately $2\times$ higher than the best code-driven baseline, reflecting a deliberate trade-off: code-driven tools optimize for cost-efficient bug detection, while \sys trades cost for richer artifact generation. The per-bug cost on RepoAudit (\$1.69/bug) remains orders of magnitude below manual audit rates (\$200--500/hour), and specification-processing phases (Phases~1--3) amortize across multiple implementations of the same specification.

\section{Extended Discussion}
\label{app:discussion}

\subsection{Unexpected results}

The most unexpected finding is the dominance of the property neighborhood effect:
11 of 15 matched issues are recovered by 2 or more distinct properties.
We initially expected a one-to-one relationship between properties and issues, but the
many-to-one structure suggests that real vulnerabilities manifest as clusters of
related property violations rather than isolated deviations.
This has implications for triage: rather than presenting individual alerts, security
tools could group findings by their shared root cause.

The false-positive phase-origin analysis revealed an unexpected compounding pattern:
most FPs originate in Phase~5, but many are seeded by weaker outputs from earlier
phases.
A property with an over-broad assertion from Phase~3, combined with a missed code
resolution in Phase~4, produces a confident-but-wrong finding in Phase~5.
This suggests that targeted improvements to early-phase output quality would
disproportionately reduce downstream false positives.

\subsection{Extended threats to validity}

\noindent\textbf{Internal validity.}
Issue matching uses LLM-assisted semantic matching with manual verification of all matches; the 4 independently discovered bugs are confirmed by fix commits, providing objective validation independent of the matching procedure.

\noindent\textbf{Construct validity.}
Precision depends on contest-organizer severity judgments. The contest's network-wide-impact threshold for Medium/High means operationally significant but localized bugs may surface as Informational or out-of-scope; our \texttt{tp} vs.\ \texttt{tp\_info} distinction preserves this transparency.

\noindent\textbf{Statistical validity.}
We do not report significance tests because the benchmarks are deterministic (single-run, fixed model configurations); RQ2 baseline numbers from RepoAudit are also single-run. The cost estimate reflects actual API charges and is reproducible at the same model pricing.


\end{document}